\documentclass[12pt]{article}
\usepackage{amsfonts}
\usepackage{amsmath}
\usepackage{amsbsy}
\usepackage{amssymb}

\font\grb=eurb10

\def\bchi{\hbox{\grb\char'037}\,}

\begin{document}
\voffset-12ex

\title{\bf Solution generating methods as ``coordinate'' transformations\\
in the solution spaces\\[-0.5ex]}
\author{G.A.~Alekseev
\footnote{e-mail: G.A.Alekseev@mi-ras.ru}\\[1ex]
\emph{V.A.Steklov Mathematical Institute}\\ \emph{Russian Academy of Sciences}
\\[-4ex]
}
\date{}

\maketitle

\begin{abstract}
The solution generating methods discovered earlier for integrable reductions of Einstein's and Einstein - Maxwell field equations (such as soliton generating techniques, B$\ddot{a}$cklund or symmetry transfor\-mations and other group-theoretical methods) can be described explicitly as transfor\-mati\-ons of especially defined ``coordi\-nates''\, in the infinite-dimensional solution spaces of these equations. In general, the role of such ``coordi\-nates'', which characterize every local solution, can be performed by the monodromy data of the fundamental solutions of the corresponding spectral problems. However for large subclasses of fields, these can be the values of the Ernst potentials on the boundaries which consist of such degenerate orbits of the space-time isometry group, in which neighbourhood the space-time geometry and electromagnetic fields possess a regular behaviour. In this paper, trans\-formations of such ``coordinates'', corresponding to diffe\-rent known solution generating procedures are described by simple enough algebraic expressions which do not need any particular choice of the initial (background) solution. Explicit forms of these transformati\-ons allow us to find the interre\-lations between the sets of free parameters, which arise in different solution generating procedures, as well as to determine some physical and geometrical properties of each generating solution even before a detail calculations of all its components.
\end{abstract}

\vfill\eject


\subsubsection*{Introduction}
Having been commenced more than forty yeas ago, the development of various approaches to studies of the internal structure of Einstein's field equations has led different authors to a discovery that \emph{in some physically important cases}, these equations are completely integrable for space-times which satisfy certain space-time symmetry conditions, called further as $\mathcal{G}_2$-symmetries.

The conditions that the space-time possesses $\mathcal{G}_2$-symmetry include the existence of two-dimensional Abelian isometry group with non-null Killing vector fields and supplementary conditions imposed on the structure of metric and matter field components. More precise descriptions of necessary symmetry properties of these fields and corresponding references can be found in \cite{SKMHH-2003}, \cite{Alekseev:2016a}. The corresponding $\mathcal{G}_2$-symmetry-reduced field equations, called here as integrable reductions of Einstein's equations, admit the solution generating procedures, i.e. various algorithms which allow, starting from arbitrarily chosen known solution of these equations, to construct infinite hierarchies of solutions with arbitrary (finite) number of free parameters.

\vspace{-2ex}
\paragraph{\rm\textit{\underline{Group-theoretic approach: Geroch group and  Kinnersley-Chitre algebra}}}\hfill\\
The development of the group-theoretic approach to the studies of internal symmetries of vacuum Einstein equations and electrovacuum Einstein - Maxwell equations had begun long  ago from beautiful discoveries of symmetry transformations for space-times with at least one Killing vector field. These transformations were found in the previous century at the end of 50th for vacuum by J.Ehlers \cite{Ehlers:1957, Ehlers:1959} and at the end of 60th for electrovacuum by B.K.Harrison \cite{Harrison:1968}. Later, it was found \cite{Kramer-Neugebauer:1969, Kinnersley:1973} (see also the book \cite{SKMHH-2003}) that these transformations represent the subgroups in a larger groups of symmetries isomorphic respectively to $SU(1,1)$ for vacuum  and to $SU(2,1)$ for electrovacuum.

At the beginning of 70th, R.Geroch published a paper \cite{Geroch:1972} where he conjectured that for vacuum space-times with two commuting non-null Killing vector fields for which two certain real constants vanish (these conditions are equivalent to the mentioned above $\mathcal{G}_2$-symmetry conditions), there exists an infinite dimensional group of internal symmetries which action on the space of solutions is transitive, i.e. every solution can be obtained applying the symmetry transformation to a chosen solution, e.g., to the Minkowski space-time.  Geroch  argued also that the corresponding infinite-dimensional algebra of infinitesimal transformations can be build inductively.
Later, in \cite{Kinnersley:1977}, W.Kinnersley extended these considerations to the case of stationary axisymmetric electrovacuum fields.\footnote{In a later paper of W.Kinnersley and D.M.Chitre \cite{KCII:1977} the authors gave an interesting comment: ``\emph{Our inclusion of electromagnetism throughout this work has been an enormous help rather than a hindrance. It has revealed a striking interrelationship between electromagnetic and gravitational fields that could not possibly have been anticipated.}''}

In the subsequent papers \cite{KCII:1977}-\cite{KCIV:1978}, W.Kinnersley and D.M.Chitre presented a systematic study of the infinite-dimensional algebra of infinitesimal symmetries of Einstein - Maxwell equations for stationary axisymmetric fields. In these papers, an infinite hierarchies of complex matrix potentials associated with every particular solution were constructed, and it was shown that these hierarchies of  potentials  form a representation space of this algebra. Moreover, it was found that for electrovacuum case, these sets of potentials admit two $3\times3$ (or $2\times 2$ in vacuum case) matrix generating functions, one of which happen to satisfy a \emph{linear system of equations with an auxiliary complex parameter}, and another generating function can be expressed algebraically in terms of the first one. Some years later, in the papers of B.Julia \cite{Julia:1983, Julia:1985} the infinite dimensional symmetry transformations of Geroch and Kinnersley and Chitre were recognized as Kac-Moody symmetries and in the paper of P.Breitenlohner and D.Maison \cite{Breitenlohner-Maison:1987} the structure of the corresponding infinite-dimensional Geroch group was described in detail. However, at that time, the problem of exponentiating of Kinnersley and Chitre infinitesinal transformations for obtaining new solutions of Einstein - Maxwell equations had remained to be solved.

\vspace{-2ex}
\paragraph{\rm\textit{\underline{Soliton solutions of Einstein and Einstein - Maxwell field equations.}}}\hfill\\
\noindent Using very different approach based on the ideas and methods of the inverse scattering theory, in the pioneer papers \cite{Belinski-Zakharov:1978, Belinski-Zakharov:1979}, V.A.Belinski and V.E.Zakharov discovered the existence of infinite hierarchies of exact $N$-soliton solutions of vacuum Einstein equations depending on $4 N$ free real parameters. These solitons can be generated on arbi\-trarily chosen vacuum background with the mentioned above $\mathcal{G}_2$-symme\-try. It is important also that in these papers the  explicit expressions had been obtained for \emph{all metric components}, including the so called conformal factor -- the coefficient in front of conformally flat part of metric. This factor was  expressed in \cite{Belinski-Zakharov:1978, Belinski-Zakharov:1979} explicitly in terms of the components of chosen vacuum background metric and solution of the corresponding spectral problem.
More compact, determinant form of Belinski and Zakharov $N$-soliton solutions was found in the author's paper \cite{Alekseev:1981}. It is worth mentioning also that in \cite{Belinski-Zakharov:1978, Belinski-Zakharov:1979}, a $2\times 2$-matrix linear singular integral equation with the kernel of a Cauchy type, was constructed for generating ``non-soliton'' vacuum solutions.

A bit later, in the author's papers \cite{Alekseev:1980a, Alekseev:1980b}, using the same general ideas and methods of the inverse scattering approach (but for essentially differennt, complex  self-dual form of Einstein-Maxwell equations found by W.Kinnersley \cite{Kinnersley:1977}), the $N$-soliton solutions of Einstein - Maxwell equations depending on $3 N$ free complex or $6 N$ real parameters were constructed starting from arbitrarily chosen ($\mathcal{G}_2$-symmetric) electrovacuum background.

\vspace{-2ex}
\paragraph{\rm\textit{\underline{B$\ddot{a}$cklund transformations.}}}
Some other solu\-tion generating methods suggested later for $\mathcal{G}_2$-symmetry-reduced vacuum Einstein equations were constructed using the basic ideas of the theory of B$\ddot{a}$cklund transformations and other features in the group-theoretic context. Namely, a construction of B$\ddot{a}$cklund transformations (for vacuum time-depen\-dent as well as for stationary axisymmetric fields) were described by Harrison \cite{Harrison:1978}\footnote{In \cite{Harrison:1978}, Harrison already mentioned Belinski and Zakharov results.}, who used the pseudopotential method of Wahlquist and Estabrook \cite{Wahlquist-Estabrook:1973}. In \cite{Harrison:1978}, the corresponding equations were expressed  in terms of closed ideal of differential 1-forms. Later, Harrison described  particular applications of  B$\ddot{a}$cklund transformations and generalized his approach to electrovacuum fields \cite{Harrison:1980,Harrison:1983}.

A bit later, Neugebauer \cite{Neugebauer:1979} presented his form of B$\ddot{a}$cklund transformations for vacuum Einstein equations for stationary axisymmetric fields, which were constructed in the spirit of the known theory of B$\ddot{a}$cklund transformations for sine-(sinh-)Gordon equation. In a short series of papers \cite{Neugebauer:1979}-\cite{Neugebauer:1980b}, Neugebauer found compact (determinant) form of $N$-fold B$\ddot{a}$cklund transformations expressed in terms of the Ernst potential of some chosen beginning (or ``initial'', or ``background'') solution and of the solutions of Riccati equations with coefficients depending on this choice.

\vspace{-2ex}
\paragraph{\rm\textit{\underline{Exponentiating of some of  Kinnersley-Chitre infinitesimal symmetries:}}}\hfill\vspace{0ex}\hskip-1ex {\rm\textit{\underline{HKX-transformations.}}}
Another solution generating method for vacuum fields had been found by C.Hoenselaers, W.Kinnersley and B.C.Xanthopoulos. This method, called later as HKX-transformations and described in \cite{HKX:1979a, HKX:1979b}, was derived as a result of exponentiation of some kinds of infinitesimal Kinnersley-Chitre transformations. The corresponding ``rank p'' transformations allow to obtain from a given initial vacuum stationary axisymmetric solution a new family of  of solutions of this type with $p+1$ arbitrary real parameters. The authors suggested also a construction of superposition of such transformations with different parameters and gave the simplest examples.

\vspace{-2ex}
\paragraph{\rm\textit{\underline{Hauser and Ernst homogeneous Hilbert problem (HHP) and their integral }}}\hfill\vspace{0ex}\hskip-1ex {\rm\textit{\underline{equation method for effecting Kinnersley-Chitre transformations.}}}\hfill\\[0ex]
Hauser and Ernst suggested yet another approach to generation of stationary axisymmetric vacuum \cite{HE:1979a}, \cite{HE:1980a}, \cite{HE:1980c}, \cite{HE:1980d} and electrovacuum \cite{HE:1979b},\cite{HE:1980b} solutions. Within the class of solutions which are regular in some neighbourhood of at least one point of the symmetry axis, the problem of ``effecting'' (i.e. exponentiating) of the Kinnersley and Chitre infinite-dimensional algebra of infinitesimal symmetry transformations was reduced to solution of a homogeneous Hilbert problem (HHP) on a closed contour on the plane of auxiliary complex parameter, which was reduced then to solution of a matrix linear singular integral equation of the Cauchy type on this contour.\footnote{As it was mentioned above, earlier a construction of $2\times 2$-matrix linear singular integral equation with the kernel of a Cauchy type, solving an appropriate Riemann-Hilbert problem on the spectral plane, was suggested for generating ``non-soliton'' vacuum solutions in the first Belinski and Zakharov paper \cite{Belinski-Zakharov:1978}. However, in \cite{HE:1979a}, the integral equation method of Hauser and Ernst was more elaborated and  examples of construction of exact solutions for rational choice of arbitrary functions in the kernal were described.} Solution of this integral equation for any chosen initial (``seed'') solution and arbitrarily selected element of Kinnersley-Chitre algebra, having been found,  allows to calculate explicitly the transformed solution. However, the suggested in \cite{HE:1979b} rational ansatz with too simple algebraic structure for solving this integral equation, had led to the class of solutions essentially more restricted in the number of free parameters than the class of electrovacuum solitons \cite{Alekseev:1980a,Alekseev:1980b}.

Two years later, using the already mentioned above pseudopotential method of Wahlquist and Estabrook \cite{Wahlquist-Estabrook:1973}, Kramer and Neugebauer \cite{Kramer-Neugebauer:1981} constructed for another set of pseudopotentials\footnote{The relations between different matrix potentials suggested for Einstein - Maxwell equations in different approaches were described by D.Kramer \cite{Kramer:1982}.} a linear system which integrability condition is also provided by Einstein - Maxwell equations for stationary axisymmetric electrovacuum fields\footnote{The structure of this system with its supplementary conditions differs essentially from more simple structure of the spectral problem  used for construction of solitons in \cite{Alekseev:1980a, Alekseev:1980b}.} and then, Neugebauer and Kramer \cite{Neugebauer-Kramer:1983} translated into the context of the system \cite{Kramer-Neugebauer:1981} a constructions of soliton solutions in the spirit of inverse scattering transform, which was used earlier in \cite{Belinski-Zakharov:1978, Belinski-Zakharov:1979} for vacuum and in \cite{Alekseev:1980a, Alekseev:1980b} for electrovacuum fields. Nonetheless, in addition to the results of  \cite{Alekseev:1980a, Alekseev:1980b},  where the calculation of all components of metric (besides only the conformal factor) and of electromagnetic potential for electrovacuum soliton solutions have been constructed\footnote{The expression for conformal factor for electrovacuum solitons was found in \cite{Alekseev:1987}}), a useful input from the paper \cite{Neugebauer-Kramer:1983} was a derivation of compact determinant expressions for the Ernst potentials for stationary axisymmetric electrovacuum solitons.

The construction of B$\ddot{a}$cklund transformations for electrovacuum Ernst equations have been  described by Harrison \cite{Harrison:1983}. An interesting feature in this paper is an application of a \emph{modified}  Wahlquist-Estabrook approach, suitable for systems of equations, which can be expressed in terms of differential forms which constitute a closed ideal with constant coefficients (CC-ideal). Many known integrable systems can be cast into such form. For these cases, it is possible to formulate simple general ansatzes which lead to a construction in some unified form of associated linear systems (and corresponding ``spectral problems'') as well as of  B$\ddot{a}$cklund transformations for these systems.

More later, a spectral problem of yet another structure for the same stationary axisymmetric electrovacuum Einstein - Maxwell field equations in the form of a sigma-model was suggested in \cite{Eris-Gurses-Karasu:1984}. In this more geometrical context, calculation of soliton solutions, also was performed in the spirit of  the inverse scattering approach, but the used form of the spectral problem has led directly to calculation of the corresponding Ernst potentials only.

\vspace{-2ex}
\paragraph{\rm\textit{\underline{On the relations between the solution generating methods.}}}\hfill\\
Close inter\-re\-la\-ti\-ons between different approaches to construction of vacuum solution generating methods (the inverse scattering method, theory of B$\ddot{a}$cklund transformations and group-theoretical approach) were described by Cosgrove \cite{Cosgrove:1980} -- \cite{Cosgrove:1982}. The relations between associated linear systems (``spectral problems'') used by different authors  for generalizations of their approaches to electrovacuum fields was found by Kramer \cite{Kramer:1982}.

\vspace{-2ex}
\paragraph{\rm\textit{\underline{On the difficulties with explicit applications of solution generating methods.}}}\hfill\\[0.3ex]
The general studies of the families of solutions generating with the methods mentioned above occur rather difficult because
these families of solutions do not admit their representation in general and explicit form, due to a presence in these solutions, besides a large number of constant parameters, some functional parameters -- the potentials which characterize the chosen initial (background) solution. In each of these  methods, these potentials should satisfy some linear systems with coefficients depending on the choice of initial solution, but this systems can be solved explicitly not for any choice of the initial solution. Only in those cases, in which for chosen initial (background) solution this linear system can be solved explicitly, one can calculate all components of the solutions generating on this background.
\vspace{-2ex}
\paragraph{\rm\textit{\underline{On the ``coordinates'' in the space of solutions.}}}
The difficulties mentioned\\ just above can be overcome if we introduce in the space of solutions, instead of metric and field components, some ``coordinates'' which, from one hand, would be related to various physical and geometrical characteristics of the solutions and, from the other hand, different solution generating procedures could be represented as transformations of these ``coordinates''.

In the most general cases, for $\mathcal{G}_2$-symmetry reduced vacuum Einstein equations and electrovacuum Einstein - Maxwell equations, the role of such ``coordinates'' in the infinite-dimensional spaces of their local solutions can belong to the monodromy data of the fundamental solutions of the corresponding associated linear systems (``spectral problems'') \cite{Alekseev:1985,Alekseev:1987}. However, for large subclasses of field configurations which possess the asymptotic behaviour of the same type near some space-time boundaries, such ``coordinates'' can be defined in a more simple way. For example, for stationary axisymmetric fields the role of such ``coordinates can be played by the values of the Ernst potentials on those parts of axis of symmetry near which the space-time geometry  and electromagnetic fields possess a regular behaviour.

In this paper, we consider the classes of fields, which possess, similarly to the regular parts of axis of symmetry in axisymmetric fields, the boundaries consisting of degenerate orbits of the isometry group $\mathcal{G}_2$ with regular behaviour of metric and electromagnetic fields near these boundaries
(see below for more details). It is clear that besides the stationary axisymmetric fields near the regular parts of the axis of symmetry, these classes of fields include, in particular, cylindrical waves and some other types of wave-like or cosmological solutions, stationary fields with Killing horizons as well as some other types of solutions which can have, in particular, a dynamical nature (like the well known ``C-metrics'').  For all these types of fields, the  ``coordinates'' in the space of solutions may be represented by the functional parameters defined as the values of the  Ernst potentials on the lines in the orbit space which consist of the points (orbits) at which the geometry of the orbits is degenerate, but the space-time geometry remains regular. These ``coordinates'' determine the corresponding local solutions ``almost unequally''.\footnote{For given local solution, the Ernst potentials are defined with some gauge freedom which does not change, however, the geometry and physical parameters of the  solution.}
\vspace{-3ex}
\paragraph{\rm\textit{\underline{Transformations of ``coordinates'' in the space of solutions.}}}\hfill\\
As it will be shown further, different solution generating procedures can be presented explicitly and in a very simple form as transformations of the described above ``coordinates'' in the spaces of local solutions. It is very important that these transformations possess a general form which does not need to specify in advance the choice of the initial solution. The initial solution in the expressions for these transformations is represented by the functions which can be chosen arbitrarily and which are the similar ``coordinates'' of the initial solution in the space of solutions.

\subsubsection*{Integrable reductions of Einstein and Einstein-Maxwell equations}
\vspace{-0.5ex}
\paragraph{\rm\textit{\underline{Metric and electromagnetic potential.}}}

Integrable reductions of vacuum Einstein equations and of electrovacuum Einstein - Maxwell equations arise if the metric and electromagnetic potential components possess the forms:
\begin{equation}\label{Components}
\begin{array}{l}
ds^2=g_{\mu\nu} dx^\mu dx^\nu+g_{ab} dx^a dx^b,\\[1.5ex]
A_i=\{A_\mu,\, A_a\},\quad A_\mu=0,
\end{array}\qquad \left\Vert\qquad
\begin{array}{l}
x^i=\{x^\mu,\,x^a\},\\[0.5ex]
  \mu,\nu,\ldots=1,2\\
  a,b,\ldots=3,4
\end{array}\right.
\end{equation}
where the components of metric  $g_{\mu\nu}$, $g_{ab}$ and of electromagnetic
potential $A_a$ are independent of the coordinates $x^a$ and may depend on  coordinates $x^\mu$.

Each of these reductions belongs to one of two types depending on
whether the space-time isometry group $\mathcal{G}_2$ admits a time-like Killing vector field (the ``elliptic'' case) or not (the ``hyperbolic'' case), i.e. whether the signature of two-dimensional metric $g_{\mu\nu}$ on the space of orbits is respectively  Euclidean or Lorentzian. In the expressions below, the sign symbol $\epsilon$ and its ``square root'' $j$ will remind us about a difference between these cases:
\begin{equation}\label{symbols}
\epsilon=\left\{\begin{array}{rcl}
1&-&\text{hyperbolic case}\\
-1&-&\text{elliptic case}
\end{array}\right. \qquad j=\left\{\begin{array}{ll}
1\,,&\epsilon=1\\
i\,,&\epsilon=-1.
\end{array}\right.
\end{equation}
The metric components $g_{\mu\nu}$ determine two-dimensional \emph{metric on the orbit space} of the space-time isometry group $\mathcal{G}_2$. By an appropriate choice of local coordinates $x^\mu$, the metric $g_{\mu\nu}$ can be presented in a conformally flat form, where we use the sign symbols $\epsilon_1$,  $\epsilon_2$ for a unified description of all cases:
\begin{equation}\label{Orbitspacemetric}
g_{\mu\nu}=f\,\eta_{\mu\nu},\qquad  \eta_{\mu\nu}=
\begin{pmatrix}\epsilon_1&0\\0&\epsilon_2\end{pmatrix},\qquad
\begin{array}{l}\epsilon_1=\pm 1,\\ \epsilon_2=\pm 1,\end{array}\qquad \epsilon_1\epsilon_2=-\epsilon.
\end{equation}
Here, by definition, $f>0$. The relation between  $\epsilon_1$, $\epsilon_2$ and $\epsilon$ arises here from the condition of the Lorentz signature $(-+++)$ of four-dimensional metric.

For the metric components $g_{ab}$, which determine the \emph{metric on the orbits} of the space-time isometry group $\mathcal{G}_2$ and for components of electromagnetic potential $A_a$, we introduce the parameterizations \begin{equation}\label{Orbitmetric}
g_{ab}=\epsilon_0\left(\begin{array}{ll}
H& H\Omega\\
H\Omega& H\Omega^2+\dfrac{\epsilon \alpha^2}{H}
\end{array}
\right),\hskip1ex A_a=\{A,\,\widetilde{A}\},\qquad
\begin{array}{l}
\det\Vert g_{ab}\Vert\equiv \epsilon\alpha^2,\\[1ex]
\end{array}
\end{equation}
where, by definition, $H>0$, $\alpha>0$ and $\epsilon_0=\pm 1$.
Then the Einstein equations as well as the Einstein - Maxwell equations for the fields (\ref{Components}) imply that the function $\alpha(x^1,x^2)$, defined in  (\ref{Orbitmetric}), is a ``harmonic'' function, i.e. it should satisfy the linear equation which is a two-dimensional d'Alembert equation in the hyperbolic case and the two-dimensional Laplace equation in the elliptic case. Therefore, for the function $\alpha(x^1,x^2)$ one can define its ``harmonically conjugated'' function $\beta(x^1,x^2)$, such that
\begin{equation}\label{AlphaBeta}
\left\{\begin{array}{l}
\det\Vert g_{ab}\Vert\equiv \epsilon\alpha^2,\\[1ex]
\eta^{\mu\nu}\partial_\mu\partial_\nu\alpha=0,
\end{array}\right.\hskip1ex
\left\{\begin{array}{l}
\partial_\mu\beta=\epsilon\varepsilon_\mu{}^\nu\partial_\nu\alpha,\\[1ex]
\eta^{\mu\nu}\partial_\mu\partial_\nu\beta=0,
\end{array}\right.\hskip1ex
\varepsilon_\mu{}^\nu=\eta_{\mu\gamma}\varepsilon^{\gamma\nu},\hskip1ex
\varepsilon^{\mu\nu}=\begin{pmatrix} 0&1\\-1&0\end{pmatrix}.
\end{equation}
These geometrically defined functions $(\alpha,\beta)$ will be used further as local conformal coordinates on the orbit space of $\mathcal{G}_2$. We call them as generalized Weyl coordinates. It is convenient also to use their linear combinations
 \begin{equation}\label{xieta}
\xi=\beta+j \alpha,\qquad \eta=\beta -j\alpha,
\end{equation}
which are real null coordinates in the hyperbolic case ($j=1$) and complex conjugated to each other in the elliptic case ($j=i$).

\paragraph{\rm\textit{\underline{Matrix form of dynamical equations for vacuum.}}}
For metrics (\ref{Components}), the dynamical part of Einstein equations for vacuum gravitational fields can be presented in  $2\times 2$-matrix form \cite{Belinski-Zakharov:1978,Belinski-Zakharov:1979}
\begin{equation}\label{g-eqn}
\left\{\begin{array}{l}
\eta^{\mu\nu}\partial_\mu(\alpha \partial_\nu \mathbf{g}\cdot \mathbf{g}^{-1})=0,\\[1ex]
\mathbf{g}^T=\mathbf{g},\quad \det\mathbf{g}\equiv
\epsilon\alpha^2,
\end{array}\right.
\end{equation}
where $\mathbf{g}=\Vert g_{ab}\Vert$, and ``${}^T$'' means the matrix transposition. Electrovacuum Einstein - Maxwell equations for the fields
(\ref{Components}) can be presented in a similar form but for $3\times 3$-matrix equations \cite{Alekseev:1987}. However there exist also more convenient forms of the dynamical part of these equations.
\vspace{-2ex}
\paragraph{\rm\textit{\underline{The Ernst equations.}}}\hskip2ex
The dynamical part of electrovacuum Einstein - Maxwell equations for space-times with $\mathcal{G}_2$-symmetry can be presented also in the form of the Ernst equations\footnote{These equations were derived originally by F.J.Ernst for stationary axisymmetric vacuum fields \cite{Ernst:1968a}, and then were generalized for the case of stationary axisymmetric electrovacuum fields \cite{Ernst:1968b}. In these equations the Weyl cylindrical coordinates were used. For these coordinates, $\alpha=\rho$, $\beta=z$. The similar equations can be derived easily in the hyperbolic case as well, and these are called usually the hyperbolic Ernst equations.} for already mentioned above function $\alpha$ and two scalar complex Ernst potentials $\mathcal{E}$ and $\Phi$. In our notations, these equations take the form (the bar on a symbol means complex conjugation):
\begin{equation}\label{Ernst-eqs}
\mathbf{\left\{\begin{array}{l}
(\hbox{Re}\,{\cal
E}+\overline{\Phi}\Phi)\,\eta^{\mu\nu}(\partial_\mu\partial_\nu
{\cal E}+\dfrac{\partial_\mu\alpha}{\alpha} \partial_\nu
{\cal E})-\eta^{\mu\nu}(\partial_\mu {\cal E}+ 2\overline{\Phi}\partial_\mu\Phi)\,\partial_\nu
{\cal E}=0,\\[1.5ex]
(\hbox{Re}\,{\cal
E}+\overline{\Phi}\Phi)\,\eta^{\mu\nu}(\partial_\mu\partial_\nu
\Phi+\dfrac{\partial_\mu\alpha}{\alpha} \partial_\nu
 \Phi)-\eta^{\mu\nu}(\partial_\mu {\cal E}+ 2\overline{\Phi}\partial_\mu\Phi)\,\partial_\nu
\Phi=0,\\[1.5ex]
\eta^{\mu\nu}\partial_\mu\partial_\nu\alpha=0.
\end{array}\right.}
\end{equation}
For $\Phi=0$, these equations reduce to vacuum Ernst equation.
The relations of $\mathcal{E}$ and $\Phi$ to the field components   (\ref{Components}) in the notations (\ref{Orbitmetric}) take the forms:
\begin{equation}\label{Comp-EF}
\left\{\!\!\begin{array}{l}
\mbox{Re}\,\Phi=A,\\
\partial_\mu(\mbox{Im}\Phi)=-\dfrac{\epsilon_0 H}{\alpha}
\varepsilon_\mu{}^\nu (\partial_\nu\widetilde{A} -\Omega \partial_\nu A),
\end{array}\right.
\left\{\!\!\begin{array}{l}
\mbox{Re}\,{\cal E}=-\epsilon_0 H-\Phi\overline{\Phi},\\
\mbox{Im}(\partial_\mu{\cal E}+2\overline{\Phi}\partial_\mu\Phi)=
\dfrac{H^2}{\alpha}\varepsilon_\mu{}^\nu\partial_\nu\Omega,
     \end{array}\right.
\end{equation}
These relations allow to find all field components (\ref{Components})  algebraically or in quadratures if the Ernst potentials $\mathcal{E}$, $\Phi$ and the function $\alpha$ are known.
\vspace{-2ex}
\paragraph{\rm\textit{\underline{The Kinnersley equations.}}}
For the field components (\ref{Components}), using the notations
\begin{equation}\label{hab-comp}
h_{ab}\equiv g_{ab},\quad h_a{}^b =
\epsilon^{bc} h_{ac}, \quad h_{ab} = h_a{}^c \epsilon_{cb},
\qquad \epsilon_{ab}=\epsilon^{ab}=\begin{pmatrix} 0 & 1 \cr -1 &
0 \end{pmatrix},
\end{equation}
the Einstein-Maxwell equations can be reduced to self-dual matrix form  \cite{Kinnersley:1977}
\begin{equation}\label{Kinnersley}
\begin{array}{l}
\begin{array}{lcl}
\!\! H_{\mu a}{}^b=i\alpha^{-1}\varepsilon_\mu{}^\nu h_a{}^c H_{\nu c}{}^b,&& H_{\mu a}{}^b=\partial_\mu H_a{}^b,\\[0.5ex]
\!\! \Phi_{\mu a}=i\alpha^{-1}\varepsilon_\mu{}^\nu h_a{}^c \Phi_{\nu c},&&
\Phi_{\mu a}=\partial_\mu \Phi_a,\\[0.5ex]
\end{array}\\[3ex]
H_{\mu a}{}^b= \partial_\mu h_a{}^b+i\alpha^{-1}\varepsilon_\mu{}^\nu h_a{}^c \partial_\nu h_c{}^b+2 \Phi_{\mu a} \overline{\Phi}{}^b
\end{array}
\end {equation}
The left equations in the first two lines of (\ref{Kinnersley}) are the self-duality conditions for $H_{\mu a}{}^b$ and $\Phi_{\mu a}$, while the existence of their potentials $H_a{}^b$ and $\Phi_a$ follow from the Einstein - Maxwell equations. The equation in the third line of (\ref{Kinnersley}) is a definition of $H_{\mu a}{}^b$, introduced in \cite{Kinnersley:1977}. The equations (\ref{Kinnersley}) were used also in \cite{Alekseev:1980a, Alekseev:1980b} in the analysis of integrability of Einstein - Maxwell equations.

\vspace{-2ex}
\paragraph{\rm\textit{\underline{The Ernst equations in terms of CC-ideal of differential forms.}}}
Other inte\-resting forms of vacuum and  electrovacuum Ernst equations were  obtained by Harrison \cite{Harrison:1978, Harrison:1983}, who used a modified Wahlquist-Estabrook formalism based on construction of a closed ideal with constant coefficients (CC-ideal) of differential 1-forms. It consists of self-dual $\eta_1,\eta_2,\eta_5,\eta_7,\eta_8$ and anti-self-dual $\eta_3,\eta_4,\eta_6,\eta_9,\eta_{10}$ forms (vacuum corresponds to $\eta_7=\eta_8=\eta_9=\eta_{10}=0$):
\[{}^\star\eta_p=\dfrac 1j \eta_p,\quad p=1,2,5,7,8\quad\text{and}\quad {}^\star\eta_p=-\dfrac 1j \eta_p,\quad p=3,4,6,9,10.
\]
Here and below, ${}^\star\! d$ means the Hodge ``star operator'': ${}^\star d\phi\equiv\varepsilon_\mu{}^\nu\partial_\nu\phi dx^\mu$).
Following \cite{Harrison:1978,Harrison:1983}), we omit farther, for simplicity, the wedge symbol $\wedge$ in the  external products of forms. If the forms $\eta_1\ldots\eta_{10}$ satisfy the conditions
\begin{equation}\label{CCidealeqs}
\begin{array}{l}
4 d\eta_1=\eta_1(\eta_4+\eta_6-\eta_3)-\eta_3\eta_5-4\epsilon_0 \eta_{10}\eta_7,\\
4 d\eta_2=\eta_2(\eta_3+\eta_6-\eta_4)-\eta_4\eta_5-4\epsilon_0 \eta_9\eta_8,\\
4 d\eta_3=\eta_3(\eta_2+\eta_5-\eta_1)-\eta_1\eta_6-4\epsilon_0 \eta_8\eta_9,\\
4 d\eta_4=\eta_4(\eta_1+\eta_5-\eta_2)-\eta_2\eta_6-4\epsilon_0 \eta_7\eta_{10},\\[1ex]
2 d\eta_5=\eta_5 \eta_6,\qquad 2 d\eta_6=-\eta_5\eta_6,\\[0.7ex]
8 d\eta_7=\eta_7(\eta_4+2\eta_6-\eta_3)+2\eta_9(\eta_1-\eta_5),\\
8 d\eta_8=\eta_8(\eta_3+2\eta_6-\eta_4)+2\eta_{10}(\eta_2-\eta_5),\\
8 d\eta_9=\eta_9(\eta_2+2\eta_5-\eta_1)+2\eta_7(\eta_3-\eta_6),\\
8 d\eta_{10}=\eta_{10}(\eta_1+2\eta_5-\eta_2)+2\eta_8(\eta_4-\eta_6),\\[0.7ex]
\eta_1\eta_2=\eta_1\eta_5=\eta_1\eta_7=\eta_1\eta_8=\eta_2\eta_5=0,\\
\eta_2\eta_7=\eta_2\eta_8=\eta_5\eta_7=\eta_5\eta_8=\eta_7\eta_8=0,\\[0.7ex]
\eta_3\eta_4=\eta_3\eta_6=\eta_3\eta_9=\eta_3\eta_{10}=\eta_4\eta_6=0,\\
\eta_4\eta_9=\eta_4\eta_{10}=\eta_6\eta_9= \eta_6\eta_{10}=\eta_9\eta_{10}=0,
\end{array}
\end{equation}
this implies the existence (locally at least) of two complex functions $\mathcal{E}(x^\mu)$,  $\Phi(x^\mu)$ and two real functions $\alpha(x^\mu)$ and $\beta(x^\mu)$, such that these forms can be expressed in terms of these functions as
\begin{equation}\label{CCidealforms}
\begin{array}{l}
\begin{array}{lclccl}
\eta_1=-\dfrac{2(\mathcal{E}_\xi+2\overline{\Phi}\Phi_\xi)}
{\epsilon_0 H}d\xi&&
\eta_2=-\dfrac{2(\overline{\mathcal{E}}_\xi+2\Phi\overline{\Phi}_\xi) }{\epsilon_0 H}d\xi&&& \eta_5=\dfrac{d\xi}{j \alpha}\\[2ex]
\eta_3=-\dfrac{2(\mathcal{E}_\eta+2\overline{\Phi}\Phi_\eta)}{\epsilon_0 H}d\eta&&
\eta_4=-\dfrac{2(\overline{\mathcal{E}}_\eta+2\Phi\overline{\Phi}_\eta) }{\epsilon_0 H}d\eta&&& \eta_6=-\dfrac{d\eta}{j \alpha}
\end{array}\\[5ex]
\eta_7=\dfrac {2\,\Phi_\xi}{\sqrt{H}} d\xi,\quad
\eta_8=\dfrac{2\,\overline{\Phi}_\xi} {\sqrt{H}}d\xi,\quad
\eta_9=\dfrac{2\Phi_\eta} {\sqrt{H}}d\eta,\quad
\eta_{10}=\dfrac{2\,\overline{\Phi}_\eta} {\sqrt{H}}d\eta,
\\[4ex]
H=-\epsilon_0(\mathcal{E}+\overline{\mathcal{E}}+ 2\Phi\overline{\Phi})
\end{array}
\end{equation}
where $\xi$ and $\eta$ were defined in terms of $\alpha$ and $\beta$ in (\ref{xieta}). Besides that, as it also follows from (\ref{CCidealeqs}), the functions $\mathcal{E}$ and $\Phi$ satisfy the Ernst equations which take in the coordinates $(\xi,\eta)$ the form:
\[\left\{\begin{array}{l}
(Re \mathcal{E}+\Phi\overline{\Phi})\bigl(2\mathcal{E}_{\xi\eta}-
\dfrac{\mathcal{E}_\xi-\mathcal{E}_\eta}{\xi-\eta}\bigr)- (\mathcal{E}_\xi+2 \overline{\Phi}\Phi_\xi)\mathcal{E}_\eta-
(\mathcal{E}_\eta+2\overline{\Phi}\Phi_\eta)\mathcal{E}_\xi=0\\[1ex]
(Re \mathcal{E}+\Phi\overline{\Phi})\bigl(2\Phi_{\xi\eta}-
\dfrac{\Phi_\xi-\Phi_\eta}{\xi-\eta}\bigr)- (\mathcal{E}_\xi+2\overline{\Phi}\Phi_\xi)\Phi_\eta-
(\mathcal{E}_\eta+2\overline{\Phi}\Phi_\eta)\Phi_\xi=0
\end{array}
\right.
\]
It is worth mention in advance that presenting of field equations in terms of closed CC-ideal of 1-forms suggests a nice way  for construction of an associated linear systems and B$\ddot{a}$cklund transformations using the ansatzes
\[d\Psi=(\sum\limits_{p=1}^{10} B^p \eta_p)\Psi,\qquad
\widetilde{\eta}_p=\sum\limits_{q=1}^{10}A_p{}^q\eta_q,
\]
where the matrix coefficients $B^p$ and $A_p{}^q$ are functions of the so called pseudopotentials (see \cite{Harrison:1978,Harrison:1983} and  the corresponding section below).

\vspace{-1ex}
\paragraph{\rm\textit{\underline{Calculation of the conformal factor.}}}
The Einstein - Maxwell equations for the field components (\ref{Components}) imply also the constraint equations which allow to calculate the conformal factor $f$ for any solution of dynamical equations (\ref{g-eqn}) or (\ref{Ernst-eqs})  (see e.g., \cite{Belinski-Zakharov:1978, Alekseev:1987, Alekseev:2016a}).
In particular, in generalized Weyl coordinates $x^1=\alpha$ and $x^2=\epsilon_1 \beta$, where the multiplier $\epsilon_1$ arises due to definition of $\beta$ given in (\ref{AlphaBeta}), we have $g_{\mu\nu}dx^\mu dx^\nu=f(\epsilon_1 d\alpha^2+\epsilon_2 d\beta^2)$ and
\begin{equation}\label{ConFactor}
 \left\{\begin{array}{l}
 \dfrac{\partial_\alpha (f H)}{f H}=\dfrac{\alpha}{2 H^2} (\mathcal{F}_\alpha \overline{\mathcal{F}}_\alpha+\epsilon
\mathcal{F}_\beta \overline{\mathcal{F}}_\beta
)-\dfrac{2\epsilon_0\alpha}{H} (\partial_\alpha\Phi \partial_\alpha\overline{\Phi}+\epsilon
\partial_\beta\Phi \partial_\beta\overline{\Phi}_\beta),\\[2ex]
 \dfrac{\partial_\beta (f H)}{f H}=\dfrac{\alpha}{2 H^2} (\mathcal{F}_\alpha \overline{\mathcal{F}}_\beta+
\overline{\mathcal{F}}_\alpha\mathcal{F}_\beta
)-\dfrac{2\epsilon_0\alpha}{H} (\partial_\alpha\Phi \partial_\beta\overline{\Phi}+
\partial_\alpha\overline{\Phi}\partial_\beta\Phi),
\end{array}\right.
\end{equation}
$\begin{array}{l}
\text{where}\quad\mathcal{F}_\alpha=-\epsilon_0\partial_\alpha H+i\alpha^{-1} H^2\partial_\beta\Omega=
\partial_\alpha\mathcal{E}+2\overline{\Phi} \partial_\alpha\Phi,\\[0.5ex]
\phantom{\text{where}\quad}\mathcal{F}_\beta=-\epsilon_0\partial_\beta H+i\epsilon \alpha^{-1} H^2\partial_\alpha\Omega=\partial_\beta\mathcal{E}+2\overline{\Phi} \partial_\beta\Phi,
\end{array}
$\\[1ex]
\noindent
and this allows to calculate the conformal factor in quadratures in terms of field components or in terms of the Ernst potentials.

\subsubsection*{Space of solutions regular near degenerate orbits
of $\mathcal{G}_2$}
The orbits of the isometry group $\mathcal{G}_2$ are degenerate if the function $\alpha$ (which characterizes the element of area on the orbits) vanishes on these orbits.
In the orbit space of the group $\mathcal{G}_2$ the points with $\alpha=0$ constitute the lines. There exists a large class of solutions with regular behaviour of space-time geometry and of electromagnetic fields on the line
$\alpha=0$ or on its finite or semi-infinite intervals. The necessary and sufficient condition of this regularity is a possibility to expand the field components (\ref{Components}) near $\alpha=0$:
\begin{equation}\label{HOmegaf-expansion}
\begin{array}{lcl}
  H=H_0+H_1 \alpha^2+H_2 \alpha^4+\ldots &&
  A=A_0+A_1 \alpha^2+A_2 \alpha^4+\ldots\\
  \Omega=\Omega_0 +\Omega_1 \alpha^2+\Omega_2 \alpha^4+\ldots &&
  \widetilde{A}=\widetilde{A}_0+\widetilde{A}_1 \alpha^2+\widetilde{A}_2 \alpha^4+\ldots\\[0.5ex]
  f=f_0+f_1 \alpha^2+f_2 \alpha^4+\ldots&&
\end{array}
\end{equation}
The corresponding expansions for the Ernst potentials possess the forms
\begin{equation}\label{Ernst-expansion}
\mathcal{E}=\mathcal{E}_0+\mathcal{E}_1 \alpha^2+\mathcal{E}_2 \alpha^4+\ldots, \qquad
\Phi=\Phi_0+\Phi_1 \alpha^2+\Phi_2 \alpha^4+\ldots
\end{equation}
where the coefficients of these expansions are functions of the generalized Weyl coordinate $\beta$ ``harmonically'' conjugated to the coordinate $\alpha$.

Substitution of the expansions (\ref{HOmegaf-expansion}) and (\ref{Ernst-expansion}) into (\ref{Ernst-eqs}), (\ref{Comp-EF}) and (\ref{ConFactor}) shows that $\Omega_0$ is constant\footnote{The constant parameter $\Omega_0$ can be made equal zero using appropriate linear transformations with constant coefficients of the Killing vectors $\partial/\partial x^a$. However, if one of the Killing vectors corresponds to an axial symmetry with $2 \pi$-periodical angle coordinate $\varphi$, this transformation of Killing vectors is not admissible global coordinate transformation and it should be considered as some ``cut-and-past'' procedure changing the space-time manifold such that the role of $2 \pi$-periodical angle coordinate will be played not by the old coordinate $\varphi$, but by some new angle coordinate $\varphi^\prime$. Besides that, it is possible that on the axis of symmetry several regular intervals separated by the sources may exist and near each of these intervals the expansions of the type  (\ref{HOmegaf-expansion}), (\ref{Ernst-expansion}) may take place. In this case, the constants $\Omega_0$ may be different on different intervals and we can not make all of them equal to zero simultaneously by any global Killing vector transformation.}; the values $\mathcal{E}_0(\beta)$ and $\Phi_0(\beta)$ of the Ernst potentials on the boundary $\alpha = 0$ remain arbitrary, while all other coefficients of these expansions are determined uniquely by these boundary values.
Thus, the space of solutions regular on the boundary $\alpha = 0$ is infinite dimensional and the functions $\mathcal{E}_0(\beta)$ and $\Phi_0(\beta)$ can serve as ``coordinates'' in this space.\footnote{Using the functions $\mathcal{E}_0(\beta)$ and $\Phi_0(\beta)$  as ``coordinates'', one has to take into account that the Ernst potential $\mathcal{E}$ is defined up to an arbitrary additive imaginary constant and the potential $\Phi$ -- up to an additive complex constant. Changes of these constants lead to  ``gauge'' transformations of $\mathcal{E}_0(\beta)$ and $\Phi_0(\beta)$ which leave the metric functions $H$ and $\Omega$ unchanged and therefore, physical properties of the solution remain unchanged.}

Further we consider the action of various known solution generating procedures in the space of solutions which Ernst potentials near $\alpha=0$  possess the expansions (\ref{Ernst-expansion}). In this case, the ``coordinates'' of generating solution $\{\mathcal{E}(\beta),\Phi(\beta)\}$ will be expressed in terms of arbitrarily chosen ``coordinates'' of the seed (background) solution $\{{\overset o {\mathcal{E}}}_0(\beta),{\overset o {\Phi}}_0(\beta)\}$ and of a set of arbitrary constants.

\subsubsection*{Inverse Scattering approach and vacuum solitons }
As it was mentioned in the Introduction, originally the soliton solutions for the Einstein equations were found for vacuum gravitational fields by Belinski and Zakharov \cite{Belinski-Zakharov:1978} for the hyperbolic case ($\epsilon=1$), and later in their paper \cite{Belinski-Zakharov:1979} for the elliptic case ($\epsilon=-1$) -- see also the book \cite{Belinski-Verdaguer:2001}. The construction of solitons suggested in  \cite{Belinski-Zakharov:1978, Belinski-Zakharov:1979} was based on a representation of the field equations (\ref{g-eqn}) as the compatibility conditions of the following linear system with a complex (``spectral'') parameter $\lambda$ for $2\times 2$-matrix function $\Psi(x^\mu,\lambda)$:
\begin{equation}\label{BZ-LinSys}
(\lambda\delta_\mu{}^\nu-\epsilon\alpha\varepsilon_\mu{}^\nu) \partial_\nu\Psi-2\lambda\beta_\mu (\partial/\partial\lambda)\Psi=V_\mu\Psi,\qquad V_\mu=-\epsilon\alpha\varepsilon_\mu{}^\nu \partial_\nu \mathbf{g}\cdot \mathbf{g}^{-1},
\end{equation}
where $\beta_\mu=\partial_\mu\beta$ and the functions $\alpha(x^\mu)$ and $\beta(x^\mu)$ were defined in (\ref{AlphaBeta}). For construction of soliton solutions for the system(\ref{BZ-LinSys}), the following ansatz  was used in  \cite{Belinski-Zakharov:1978, Belinski-Zakharov:1979} for $\Psi(x^\mu,\lambda)$:
\begin{equation}\label{BZdressing}
\Psi=\bchi\cdot{\overset \circ {\Psi}},\qquad
\bchi=\mathbf{I}+\mathop{\sum}_{k=1}^{N} \dfrac{R_k}{\lambda-\mu_k},\quad
\bchi^{-1}=\mathbf{I}+\mathop{\sum}_{\ell=1}^{N} \dfrac{S_\ell}{\lambda-\nu_\ell},
\end{equation}
where ${\overset \circ {\Psi}}(x^\mu,\lambda)$ is a fundamental solution of the system (\ref{BZ-LinSys}), corresponding to some known vacuum solution chosen  as the background for solitons. (Here and below, the symbol ``${}^o$'' on the letter denotes all functions characterizing this choice of the background solution).

The functions   $\mu_k$ and $\nu_\ell$ as well as the matrices $R_k$ and $S_\ell$ in (\ref{BZdressing}) are independent of $\lambda$ and these are unknown functions of $x^\mu$. Substitution of (\ref{BZdressing}) into the equations (\ref{BZ-LinSys}) and subsequent solution of the relations, derived in this way, allow to calculate all metric components and the Ernst potential for the solution which describe $N$ solitons on the chosen vacuum background.

\vspace{-2ex}
\paragraph{\rm\textit{\underline{Determinant form of vacuum $N$-soliton solutions}}}
All metric components and the Ernst potential for Belinski and Zakharov vacuum $N$-soliton solutions can be expressed in a more compact (determinant) form which was found by the author in \cite{Alekseev:1981}. Here we present only the general determinant expression for the Ernst potential of $N$-soliton solution used below:
\begin{equation}\label{ErnstEbz}\text{Re\,}\mathcal{E}=(\mathop\Pi_{k=1}^{N}\lambda_k) \left(\dfrac{\Delta_{\scriptscriptstyle{[Re]}}}\Delta\right) \text{Re\,}{\overset \circ{\mathcal{E}}},\qquad
\text{Im\,}\mathcal{E}= \text{Im\,}{\overset \circ{\mathcal{E}}}- \dfrac{\Delta_{\scriptscriptstyle{[Im]}}}\Delta
\end{equation}
In these expressions, $N$  means a number of solitons, which is equal to a number of poles of ``dressing'' matrix function $\bchi$ and  which should be even in the case of Belinski and Zakharov solitons\footnote{In the case of odd number of solitons, in the elliptic case the Belinski and Zakharov soliton generating procedure leads to the solutions which metric signature is changed in comparison with that for initial solution, and in the hyperbolic case, the corresponding generated solutions describe the waves which possess singularities on the null wave fronts.}. In (\ref{ErnstEbz}), $\Delta$, $\Delta_{[Re]}$ and $\Delta_{[Im]}$  denote the determinants of $N\times N$-matrices ($i,j,\ldots=1,\ldots, N$):
\begin{equation}\label{Deltasbz}\Delta_{ij}=\dfrac{\lambda_i\lambda_j} {\lambda_i\lambda_j-\epsilon} (\mathbf{m}_i\cdot \mathbf{p}_j),\quad
\begin{array}{l}
\Delta_{\scriptscriptstyle{[Re]}ij}=\Delta_{ij}-({\text{Re\,}{\overset \circ{\mathcal{E}}}})^{-1} (\mathbf{p}_i\cdot \mathbf{e}_1)(\mathbf{p}_j\cdot \mathbf{e}_1),\\[1ex]
\Delta_{\scriptscriptstyle{[Im]}ij}=\Delta_{ij}+\mu_j (\mathbf{p}_i\cdot \mathbf{e}_1)(\mathbf{m}_j\cdot \mathbf{e}_2),
\end{array}
\end{equation}
where $(\mathbf{m}_i\cdot \mathbf{p}_j)$ means the scalar products of vectors $\mathbf{m}_i$ and $\mathbf{p}_j$, while the vectors $\mathbf{e}_1$ and $\mathbf{e}_2$ possess the components $\mathbf{e}_1=\{1,\,0\}$ and $\mathbf{e}_2=\{0,\,1\}$. The functions $\lambda_k=\mu_k/\alpha$, and  $\mu_k(\alpha,\beta)$ are solutions of the following algebraic equations, while the components of vectors $\mathbf{m}_k$ and $\mathbf{p}_k$ possess the expressions:
\begin{equation}\label{MuNm}
\mu_k+2\beta+\dfrac{\epsilon\alpha^2} {\mu_k}=2 w_k,\qquad
\mathbf{p}_k={\overset \circ{\mathbf{g}}}\cdot  \mathbf{m}_k^T,\qquad
\mathbf{m}_k=\{1,\,c_k\}\cdot {\overset \circ{\mathbf{M}}}{}_k .
\end{equation}
In these expressions, $w_k$ and $c_k$ are constants which can be chosen arbitrarily, provided the sets $w_k$, $c_k$, as well as the corresponding functions $\mu_k$ should consist of pairs of real and/or complex conjugated to each other functions. The $2\times 2$-matrix ${\overset \circ{\mathbf{g}}}$ consists of components ${\overset \circ{g}}_{ab}$ of the chosen background metric and each matrix ${\overset \circ{\mathbf{M}}}{}_k$ is a fundamental solution of a linear system
\begin{equation}\label{Mk-eqs}
\bigl(\mu_k\delta_\mu{}^\nu- \epsilon\alpha\varepsilon_\mu{}^\nu\bigr)\partial_\nu
{\overset \circ{\mathbf{M}}}{}_k +{\overset \circ{\mathbf{M}}}{}_k {\overset \circ {\mathbf{V}}}{}_\mu=0,\quad\text{where}\quad {\overset \circ {\mathbf{V}}}{}_\mu=-\epsilon\alpha\varepsilon_\mu{}^\nu\partial_\nu
{\overset \circ {\mathbf{g}}}\cdot {\overset \circ {\mathbf{g}}}{}^{-1}
\end{equation}
The most difficult point in these calculations is finding of the fundamental solution ${\overset \circ{\mathbf{M}}}{}_k$ of the equations (\ref{Mk-eqs}).
However, this system can be solved explicitly not for any chosen background solution ${\overset \circ {\mathbf{g}}}$, and this does not allow to express $N$-soliton solution explicitly in terms of the components of the background metric and a set of arbitrary constant parameters. And, even in the explicitly solvable cases, a possible complexity of the derived expression can make useful a consideration of asymptotic behaviour of the solutions.

\vspace{-1ex}
\paragraph{\rm\textit{\underline{Asymptotics of vacuum $N$-soliton solution near $\alpha=0$.}}}\quad\hfill\\[0.5ex]
Instead of specification of the choice of the initial (background) solution, we assume only that near $\alpha=0$  its asymptotic possess the form (\ref{HOmegaf-expansion}) - (\ref{Ernst-expansion}). Besides that, we can restrict our consideration by the minimal case $N=2$. because, as it was shown by Belinski and Zakharov \cite{Belinski-Zakharov:1979}, the generating solution procedure can be constructed iteratively, i.e. the $N$-soliton solution can be constructed generating two solitons on the background plus $N-2$ solitons.

For any $w_k$ (real or complex) the functions $\mu_k(\alpha,\beta)$ for $\alpha\to 0$, can have one of two possible asymptotics which we  denote by the indices ``${}^+$'' or ``${}^-$'':
\[\begin{array}{l}
\mu_k^{\scriptscriptstyle{+}}=\dfrac{\epsilon\alpha^2} {2(w_k^{\scriptscriptstyle{+}}-\beta)}\bigl[1+ \dfrac{\epsilon\alpha^2} {4(w_k^{\scriptscriptstyle{+}}-\beta)^2}+\ldots\bigr],\\[2ex]
\mu_k^{\scriptscriptstyle{-}}={2(w_k^{\scriptscriptstyle{-}}-\beta)}\bigl[1- \dfrac{\epsilon\alpha^2} {4(w_k^{\scriptscriptstyle{-}}-\beta)^2}+\ldots\bigr].
\end{array}
\]
Therefore, the calculations using the general expressions (\ref{ErnstEbz})--(\ref{Mk-eqs}) should be divided into the subcases and even in the case $N=2$ we have to consider separately three cases in which both $\mu_1$ and $\mu_2$ possess the same type of asymptotics ($\mu_1^+$, $\mu_2^+$) or ($\mu_1^-$, $\mu_2^-$), or a ``mixed'' case ($\mu_1^+,\mu_2^-$)  can take place. Then, in the zero-order terms with respect to $\alpha$ we obtain the corresponding boundary values of the Ernst potential $\mathcal{E}_{++}(\beta)$, $\mathcal{E}_{--}(\beta)$ and $\mathcal{E}_{+-}(\beta)$ for vacuum two-soliton solution on vacuum background described by an arbitrary boundary value of its Ernst potential ${\overset \circ{\mathcal{E}}}(\beta)$:
\[\begin{array}{l}
\mathcal{E}_{\scriptscriptstyle{++}}(\beta)=\dfrac{i\left[ c_2^{\scriptscriptstyle{+}}(w_1^{\scriptscriptstyle{+}}-\beta) - c_1^{\scriptscriptstyle{+}}(w_2^{\scriptscriptstyle{+}}-\beta)\right]
{\overset \circ{\mathcal{E}}}(\beta)+ w_1^{\scriptscriptstyle{+}}-w_2^{\scriptscriptstyle{+}}} {c_1^{\scriptscriptstyle{+}} c_2^{\scriptscriptstyle{+}}(w_1^{\scriptscriptstyle{+}} -w_2^{\scriptscriptstyle{+}}){\overset \circ{\mathcal{E}}}(\beta)-i c_1^{\scriptscriptstyle{+}}(w_1^{\scriptscriptstyle{+}}-\beta) +i c_2^{\scriptscriptstyle{+}}(w_2^{\scriptscriptstyle{+}}-\beta)}\\[3ex]
\mathcal{E}_{\scriptscriptstyle{--}}(\beta)=\dfrac{-i\left[ c_1^{\scriptscriptstyle{-}}(w_1^{\scriptscriptstyle{-}}-\beta) - c_2^{\scriptscriptstyle{-}}(w_2^{\scriptscriptstyle{-}}-\beta)\right]
{\overset \circ{\mathcal{E}}}(\beta)+ c_1^{\scriptscriptstyle{-}}c_2^{\scriptscriptstyle{-}}(w_1^{\scriptscriptstyle{-}} -w_2^{\scriptscriptstyle{-}})}
{(w_1^{\scriptscriptstyle{-}} -w_2^{\scriptscriptstyle{-}}){\overset \circ{\mathcal{E}}}(\beta)-i c_1^{\scriptscriptstyle{-}}(w_2^{\scriptscriptstyle{-}}-\beta) +i c_2^{\scriptscriptstyle{-}}(w_1^{\scriptscriptstyle{-}}-\beta)}\\[3ex]
\mathcal{E}_{\scriptscriptstyle{+-}}(\beta)= \dfrac{[w_{\scriptscriptstyle{+}}-\beta +  c_{\scriptscriptstyle{+}}  c_{\scriptscriptstyle{-}}(w_{\scriptscriptstyle{-}}-\beta)] {\overset \circ{\mathcal{E}}}(\beta)+ i c_{\scriptscriptstyle{-}}(w_{\scriptscriptstyle{+}}-w_{\scriptscriptstyle{-}})} {w_{\scriptscriptstyle{-}}-\beta +  c_{\scriptscriptstyle{+}}  c_{\scriptscriptstyle{-}}(w_{\scriptscriptstyle{+}}-\beta) - i c_{\scriptscriptstyle{+}}(w_{\scriptscriptstyle{+}}-w_{\scriptscriptstyle{-}})
{\overset \circ{\mathcal{E}}}(\beta)}
\end{array}
\]
It is necessary to note here that for $N=2$ in the mixed case $(+-)$, the constants $w_+$, $w_-$, as well as the constants $c_+$, $c_-$ should be real, while in the cases $(++)$ and $(--)$ the pairs of constants  $w_1^+$, $w_2^+$, and $c_1^+$, $c_2^+$, as well as $w_1^-$, $w_2^-$, and $c_1^-$, $c_2^-$ can be chosen real or complex conjugated to each other. However, it is easy to see that the solutions with boundary values of the Ernst potential  $\mathcal{E}_{\scriptscriptstyle{++}}(\beta)$ and $\mathcal{E}_{\scriptscriptstyle{--}}(\beta)$ are not different and these can be transformed one to another by a substitution of the parameters
\[w_1^{\scriptscriptstyle{-}}=w_1^{\scriptscriptstyle{+}},\quad
w_2^{\scriptscriptstyle{-}}=w_2^{\scriptscriptstyle{+}},\quad
c_1^{\scriptscriptstyle{-}}=-1/{c_1^{\scriptscriptstyle{+}}},\quad c_2^{\scriptscriptstyle{-}}=-1/{c_2^{\scriptscriptstyle{+}}}.
\]
 The two-soliton solutions of the mixed type with the boundary value of the Ernst potential $\mathcal{E}_{\scriptscriptstyle{+-}}(\beta)$ occur to be a part of the family $\mathcal{E}_{\scriptscriptstyle{++}}$ which corresponds to real parameters $w_1^{\scriptscriptstyle{+}}$,
$w_2^{\scriptscriptstyle{+}}$ and  $c_1^{\scriptscriptstyle{+}}$,
$c_2^{\scriptscriptstyle{+}}$. To see this, it is enough to substiute
\[w_1^{\scriptscriptstyle{+}}=w_{\scriptscriptstyle{+}},\quad
w_2^{\scriptscriptstyle{+}}=w_{\scriptscriptstyle{-}},\quad
c_1^{\scriptscriptstyle{+}}=c_{\scriptscriptstyle{+}},\quad
c_2^{\scriptscriptstyle{+}}=-1/c_{\scriptscriptstyle{-}}.
\]
Thus, the most general vacuum two-soliton solution generated on an arbitrary vacuum background with metric of the form (\ref{Components}) is determined by the Ernst potential with the boundary value at $\alpha=0$ of the form
\begin{equation}\label{EotoE}
\mathcal{E}(\beta)=\dfrac{i\left[ c_2 (w_1-\beta) - c_1(w_2-\beta)\right]
{\overset \circ{\mathcal{E}}}(\beta)+ w_1-w_2} {c_1 c_2(w_1 -w_2)\,{\overset \circ{\mathcal{E}}}(\beta)-i c_1(w_1-\beta) +i c_2(w_2-\beta)},
\end{equation}
where ${\overset \circ{\mathcal{E}}}(\beta)$ is a boundary value at $\alpha=0$ of the Ernst potential of the background metric; $(w_1,w_2)$ and $(c_1,c_2)$ are the pairs of arbitrary chosen constant parameters, which should be real or complex conjugated to each other.

We note also that the cases of the choice of parameters in the pairs $(w_1,w_2)$ and $(c_1,c_2)$  as real or complex conjugated to each other are essentially different. In particular, in stationary axisymmetric case with real $w_k$ and $c_k$ the two-soliton solution describes the interaction of a rotating black hole (Kerr-NUT source with a horizon) with the background field, and the choice of these parameters complex conjugated in pairs leads to the solution which describes the interaction of a naked singularity without a horizon (``overextreme'' Kerr-NUT source) with the same background.

Subsequent applications of Belinski and Zakharov vacuum soliton generating transformations lead to solutions with any even number of solitons on a chosen background. The corresponding boundary values of these soliton solutions can be derived easily by a subsequent transformations of the type (\ref{EotoE}) for the boundary data for the Ernst potential on $\alpha=0$.

\subsubsection*{Inverse Scattering approach and Einstein - Maxwell solitons }
Later attemps of a direct generalization of the soliton generating transformations from vacuum to electrovacuum fields had not been successful. Another method for generating electrovacuum solitons on arbitrarily chosen electrovacuum background was suggested in the author's papers \cite{Alekseev:1980a, Alekseev:1980b} (see \cite{Alekseev:1987} for more details). This method is based on a complex form of the Einstein - Maxwell equations expressed in the form of duality equations. These equations were presented as integrability conditions of a linear system with a constant (``spectral'') parameter $w$ for $3\times 3$-matrix function $\Psi(x^\mu,w)$, which was supplied with the condition of existence  of a matrix integral $\mathbf{K}(w)$ of Hermitian structure:
\begin{equation}\label{SD-LinSys}
2 i[(w-\beta)\delta_\mu{}^\nu-\epsilon\alpha\varepsilon_\mu{}^\nu] \partial_\nu \Psi=\mathbf{U}_\mu \Psi, \qquad  \Psi^\dagger\mathbf{W}\Psi=\mathbf{K}(w),
\end{equation}
where ``${}^\dagger$'' means the Hermitian conjugation. In vacuum limit, this system does not coincide with the linear system of Belinski and Zakharov  \cite{Belinski-Zakharov:1978, Belinski-Zakharov:1979}.
The complex $3\times 3$-matrices $\mathbf{U}_\mu$ ($\mu=1,2$) are independent of the complex parameter $w$ and their components can be expressed  in terms of the metric and electromagnetic potential components and their first derivatives. The components of the Hermitian $3\times 3$-matrix $\mathbf{W}$ can be expressed algebraically in terms of metric components $g_{ab}$ and components of complex electromagnetic potential $\Phi_a$. In contrast to \cite{Alekseev:1980a, Alekseev:1980b} and \cite{Alekseev:1987}, the explicit expressions for $3\times 3$-matrices $\mathbf{U}_\mu$  and $\mathbf{W}$ are given below for the case of ``maximally positive'' metric signature $(-+++)$ used here (other notations see in (\ref{hab-comp}) and (\ref{Kinnersley})):
\[
\mathbf{U}_\mu=\left(\begin{array}{ll}
-H_{\mu a}{}^b&\Phi_{\mu a}\\[1ex]
-2\overline{\Phi}{}^c H_{\mu c}{}^b&2\overline{\Phi}{}^c\Phi_{\mu c}
\end{array}\right),\hskip1ex \mathbf{W}=4 i(w-\beta)\mathbf{\Omega} +\begin{pmatrix}
4 h^{ab}+4\Phi^a\overline{\Phi}{}^b&-2 \Phi^a\\[1ex]
-2\overline{\Phi}{}^b&1
\end{pmatrix},
\]
where $\mu=1,2$ and $a,b,\ldots=3,4$. The matrix $\mathbf{\Omega}$ and  the reduced form of the integral $\mathbf{K}(w)$ possess the expressions
\begin{equation}\label{WandK}
\mathbf{\Omega}=\begin{pmatrix}
0&\! 1&\! 0\\
-1&\! 0&\! 0\\
0&\! 0&\! 0
\end{pmatrix},\qquad
\mathbf{K}(w)=\begin{pmatrix}
0&\! 4 i&\! 0\\
-4 i&\! 0&\! 0\\
0&\! 0&\! 1
\end{pmatrix}.
\end{equation}
The soliton solutions of Einstein - Maxwell equations arise as the solutions of the spectral problem (\ref{SD-LinSys}) with the ansatz similar to that mentioned  in (\ref{BZdressing}):
\begin{equation}\label{SDdressing}
\Psi=\bchi\cdot{\overset \circ {\Psi}},\qquad
\bchi=\mathbf{I}+\mathop{\sum}_{k=1}^{N} \dfrac{R_k}{w-w_k},\quad
\bchi^{-1}=\mathbf{I}+\mathop{\sum}_{\ell=1}^{N} \dfrac{S_\ell}{w-\overline{w}_\ell},
\end{equation}
where $w_k$ are arbitrary complex constants, the $3\times 3$-matrix functions $R_k(x^\mu)$ and $S_l(x^\mu)$ are the unknowns, which are independent of $w$, and ${\overset \circ {\Psi}}(x^\mu,w)$ is a fundamental solution of the system (\ref{SD-LinSys}), corresponding to arbitarily chosen ($\mathcal{G}_2$-symmetric) electrovacuum background for solitons.

In accordance with \cite{Alekseev:1980a, Alekseev:1980b} the expressions for the Ernst potentials for $N$-soliton solution of electrovacuum Einstein - Maxwell equations can be presented in the forms
\begin{equation}
\begin{array}{l}
\mathcal{E}={\overset \circ {\mathcal{E}}}-2 i \sum\limits_{k,\ell=1}^{N} \Delta^{-1}_{k\ell}(e_1\cdot\mathbf{p}_\ell)(\mathbf{m}_k\cdot e_2),\\[1ex]
\Phi={\overset \circ {\Phi}}+2 i \sum\limits_{k,\ell=1}^{N} \Delta^{-1}_{k\ell}(e_1\cdot\mathbf{p}_\ell)(\mathbf{m}_k\cdot e_3),
\end{array}
\end{equation}
where ${\overset \circ {\mathcal{E}}}$ and ${\overset \circ {\Phi}}$ are the potentials of the chosen background solution; $(e_1\cdot\mathbf{p}_\ell)$, $(\mathbf{m}_k\cdot e_2)$ and $(\mathbf{m}_k\cdot e_3)$ are respectively the first component of each of the vectors $\mathbf{p}_\ell$, the second and third components of each of the vectors $\mathbf{m}_k$ (the indices  $k,\ell=1,2,\ldots, N$ numerate not the components, but the three-dimensional vectors $\mathbf{p}_\ell$ and $\mathbf{m}_k$). The components of the matrix $\Vert\Delta_{k\ell}\Vert$ are:
\[\Delta_{k\ell}=\dfrac{(\mathbf{m}_\ell\cdot\mathbf{p}_k)} {w_\ell-\overline{w}_k},\qquad w_k\ne\overline{w}_k,\qquad k,\ell=1,2,\ldots, N,
\]
where $\{w_k\}$  is a set of $N$ arbitrary complex constants which should not be chosen real;  $(\mathbf{m}_\ell\cdot\mathbf{p}_k)$ is a scalar product of vectors $\mathbf{m}_\ell$ and $\mathbf{p}_k$. The components of vectors $\mathbf{m}_\ell$ and $\mathbf{p}_k$ can be expressed in terms of the fundamental solution ${\overset \circ {\Psi}}(x^\mu,w)$ of the system  (\ref{SD-LinSys}) and the value of its matrix integral ${\overset \circ {\mathbf{K}}}(w)$, corresponding to the chosen background solution in the forms
\begin{equation}\label{Parameters}
\begin{array}{lcl}
\mathbf{m}_\ell=\mathbf{k}_\ell \cdot {\overset \circ {\Psi}}{}^{-1}(x^\mu,w=w_\ell),&&\mathbf{k}_\ell=\{1,
c_\ell,\, d_\ell\},\\[0ex]
\mathbf{p}_k= {\overset \circ {\Psi}}(x^\mu,w=\overline{w}_k) \cdot\mathbf{l}_k,&&\mathbf{l}_k=4 i {\overset \circ {\mathbf{K}}}{}^{-1}(\overline{w}_k)\cdot \mathbf{k}_k^\dagger,
\end{array}\quad \mathbf{l}_k=\begin{pmatrix}
-\overline{c}_k\\1\\ 4i\overline{d}{}_k\end{pmatrix},
\end{equation}
where $\{c_\ell\}$ and $\{d_k\}$ are two sets each consisting of $N$ arbitrary complex constants, and the last expression for the components of the vectors  $\mathbf{l}_k$ corresponds to the choice of ${\overset \circ {\mathbf{K}}}(w)$ in the form (\ref{WandK}). The $N$-soliton electrovacuum solutions described above arise as the result of soliton generating transformations of arbitrarily chosen electrovacuum solution possessing $\mathcal{G}_2$-symmetry, which plays the role of background for solitons. These solutions depend on $3 N$ arbitrary complex constants $c_k$, $d_k$ and $w_k$.

\vspace{-1ex}
\paragraph{\rm\textit{\underline{Asymptotics of Einstein - Maxwell $N$-soliton solutions near $\alpha=0$.}}}\quad\hfill\\[0.5ex]
The system (\ref{SD-LinSys}), as well as the system (\ref{Mk-eqs}), does not admit an explicit solution in general form without a particular choice of the background solution. Meanwhile, for any solution with regular behaviour of gravitational and electromagnetic fields for $\alpha\to 0$, there exists a fundamental solution of the system (\ref{SD-LinSys}), which boundary value at $\alpha=0$ can be presented in the form
\[{\overset \circ {\Psi}}(\beta,w)=\begin{pmatrix}
\dfrac 1{w-\beta}&-\dfrac{\overset\circ{\mathcal{E}}(\beta)} {2 i(w-\beta)}&\dfrac{\overset\circ{\Phi}(\beta)} {2 i(w-\beta)}\\
0&1&0\\
0&0&1
\end{pmatrix},
\]
where ${\overset \circ {\mathcal{E}}(\beta)}$ and  ${\overset\circ{\Phi}(\beta)}$ are the boundary values at $\alpha=0$ of the Ernst potentials of background solution arbitrarily chosen within the class of solutions which are regular near $\alpha=0$. In the case $N=1$, using  electrovacuum soliton generating  transformation, we obtain in the leading term for $\alpha\to 0$ the relation between the boundary values of the Ernst potentials for the background and generating one-soliton solution on this background
\begin{equation}\label{EoFotoEF}
\begin{array}{l}
\mathcal{E}=\dfrac{4 c (w_1-\overline{w}_1)(\overline{c}-2 \overline{d}{\overset\circ{\Phi}})+2 i {\overset\circ{\mathcal{E}}}[\overline{c}(\overline{w}_1-\beta)-c(w_1-\beta)-4 i d \overline{d}(\overline{w}_1-\beta)]}
{(w_1-\overline{w}_1)({\overset\circ{\mathcal{E}}}-4 i \overline{d}{\overset\circ{\Phi}})+ 2 i [\overline{c}(w_1-\beta)-c(\overline{w}_1-\beta)-4 i d \overline{d}(\overline{w}_1-\beta)]}\\[3ex]
\Phi=
\dfrac{2 i d(w_1-\overline{w}_1)({\overset\circ{\mathcal{E}}}
+2 i \overline{c})-2 i{\overset\circ{\Phi}} [(c-\overline{c})(\overline{w}_1-\beta)+
4 i d \overline{d}(w_1-\beta)]}
{(w_1-\overline{w}_1)
({\overset\circ{\mathcal{E}}}-4 i \overline{d}{\overset\circ{\Phi}})+ 2 i [\overline{c}(w_1-\beta)-c(\overline{w}_1-\beta)-4 i d \overline{d}(\overline{w}_1-\beta)]
}
\end{array}
\end{equation}
where, in accordance with (\ref{Parameters}), $c=c_1$ and $d=d_1$. It is interesting to note that the transformation (\ref{EoFotoEF}) of boundary values of the Ernst potentials possesses an important property that if the pole becomes real and the other parameters satisfy the condition $c-\overline{c}+4 i d \overline{d}\ne 0$, the soliton disappear, i.e.
\[\mathcal{E}\to {\overset\circ{\mathcal{E}}}\quad\text{and}\quad \Phi\to {\overset\circ{\Phi}} \quad\text{with}\quad w_1\to\overline{w}_1.
\]
If we put in (\ref{EoFotoEF}) ${\overset \circ {\Phi}}=0$ and $d=0$, we obtain $\Phi = 0$, i.e. the transformation (\ref{EoFotoEF}) reduces to pure vacuum soliton generating transformation. It is easy to see that
the one-soliton ($N=1$) generating transformation (\ref{EoFotoEF}) generalizes that part of  Belinski and Zakharov vacuum two-soliton ($N=2$) generating transformation (\ref{EotoE}), which corresponds to the pairs of complex conjugated poles $w_2=\overline{w}_1$ and constants $c_2=\overline{c}_1$. The relation between the parameters of these transformations are  $c_1=-1/2 \overline{c}$ and $c_2=-1/2 c$.\footnote{It seems useful to clarify here that the number of solitons means the number of simple poles in the dressing matrix $\bchi$ on the spectral plane $\lambda$ in the case of Belinski and Zakharov solitons  (\ref{BZdressing}) and on the spectral plane $w$ in the case of electrovacuum solitons (\ref{SDdressing}). Because of an obvious difference between these two techniques and of the structures of the ``spectral'' planes $\lambda$ and $w$, vacuum part of solutions with $N$ electrovacuum  solitons
should be compared with solutions with $2N$ Belinski and Zakharov vacuum solitons. This comparison shows that in contrast to Belinski-Zakharov vacuum solitons, the number $N$ of solitons (i.e. poles) in
(\ref{SDdressing}) can be not only even, but it can be odd as well. Vacuum restriction of electrovacuum $N$-soliton soluion (\ref{SDdressing})  coincides with vacuum Belinski-Zakharov  $2 N$-soliton solution with complex conjugated pairs of poles, while the electrovacuum generalization of Belinski and Zakharov vacuum solitons with pairs of real poles does not arise in this way. However, the electrovacuum solutions of soliton type with real poles may arise (at least  for special choices of the background solutions) as a result of analytical continuations of electrovacuum soliton solutions with complex poles in the space of their constant parameters.}

\subsubsection*{B$\ddot{a}$cklund transformations for vacuum and electrovacuum fields}
Other solution generating methods for vacuum and electrovacuum fields with the components  (\ref{Components}) were based  on the theory of B$\ddot{a}$cklund transformations. First of these were found by Harrison in \cite{Harrison:1978}, where both the hyperbolic and the elliptic cases of vacuum Ernst equation were considered. The procedure of construction of  B$\ddot{a}$cklund transformations suggested in \cite{Harrison:1978} had not been presented by the author in some final form, and some examples he considered later in \cite{Harrison:1980}.
The B$\ddot{a}$cklund transformations for Einstein - Maxwell equations were constrcuted by this author a few years later, in \cite{Harrison:1983}.

\vspace{-1ex}
\paragraph{\bf\textit{\underline{Harrison B$\ddot{a}$cklund transformations for vacuum fields.}}}\quad\hfill\\[1ex]
In the paper \cite{Harrison:1978}, Harrison  used the Wahlquist-Estabrook pseudipotential method \cite{Wahlquist-Estabrook:1973}. Later, he described these transformations using the modified Wahlquist-Estabrook method \cite{Harrison:1983}, which was based on the field equations in the form of a closed ideal of 1-forms with constant coefficients (CC-ideal). This form of the field equations was described above in the equations (\ref{CCidealeqs}) and (\ref{CCidealforms}), in which for vacuum case, we should set $\Phi=0$ and therefore, $\eta_7=\eta_8=\eta_9=\eta_{10}=0$. For vacuum case, 1-forms constituting CC-ideal are
\[
\eta_1=\dfrac{\mathcal{E}^\prime d\xi}{\text{Re}\mathcal{E}},\hskip1ex
\eta_2=\dfrac{\overline{\mathcal{E}}^\prime d\xi}{\text{Re}\mathcal{E}},\hskip1ex
\eta_3=\dfrac{\overline{\mathcal{E}}^\prime d\eta}{\text{Re}\mathcal{E}},\hskip1ex
\eta_4=\dfrac{\mathcal{E}^\prime d\eta}{\text{Re}\mathcal{E}},\quad
\eta_5=\dfrac{d\xi}{j \alpha},\hskip1ex\eta_6=-\dfrac{d\eta}{j \alpha}
\]
and the equation for pseudopotential $q$ found by Harrison \cite{Harrison:1978} takes the form
\[4 dq=(1+q\zeta)(q\eta_1-\dfrac 1{\zeta}\eta_3)+(q+\zeta) (-\eta_2+\dfrac{q}{\zeta}\eta_4)+
(1-q^2)(\zeta \eta_5+\dfrac{1}{\zeta}\eta_6),
\]
where $\zeta=\sqrt{(w-\eta)/(w-\xi)}$ with $w$ as an arbitrary real constant. For Harrison B$\ddot{a}$cklund transformations \cite{Harrison:1978}, the transformed 1-forms are
\[\begin{array}{lcl}
\widetilde{\eta}_1=-\dfrac{q(1+q\zeta)}{q+\zeta}\eta_1+(1+q\zeta)\eta_5,&&
\widetilde{\eta}_2=-\dfrac{q+\zeta}{q(1+q\zeta)}\eta_2+\dfrac{q+\zeta}q \eta_5,\\
\widetilde{\eta}_3=-\dfrac{(1+q\zeta)}{q(q+\zeta)}\eta_3+\dfrac{(1+q\zeta)} {q\zeta}\eta_6,&&
\widetilde{\eta}_4=-\dfrac{q(q+\zeta)}{1+q\zeta}\eta_4+\dfrac{(q+\zeta)} {\zeta}\eta_6,\\
\end{array}
\]

\vspace{-1ex}
\paragraph{\rm\textit{\underline{Asymptotics of Harrison's vacuum B$\ddot{a}$cklund transformations near $\alpha=0$.}}}\quad\hfill\\[0.5ex]
For vacuum fields which are regular near degenerate orbits with $\alpha=0$ and which admit the expansions (\ref{HOmegaf-expansion}), (\ref{Ernst-expansion}), one can obtain from the above expressions the following expansions for $\zeta$ and for pseudopotential $q$:
\[\zeta=1+\dfrac{j\alpha}{w-\beta}+\dfrac{\epsilon\alpha^2}{2(w-\beta)^2}+\ldots, \quad q=-1+q_1 \alpha+q_2 \alpha^2+\ldots,
\]
where the coefficients $q_1=\dfrac{j(\mathcal{E}_0-\overline{\mathcal{E}}_0+2 i k_0)}{(\mathcal{E}_0+\overline{\mathcal{E}}_0)(w_0-\beta)}$ and $q_2=-\dfrac 12 q_1^2$. The function $\mathcal{E}_0(\beta)$ is the value of the Ernst potential on the boundary $\alpha=0$ for the solution chosen for application of the B$\ddot{a}$cklund transformation. The expressions given just above lead to the following expressions for the boundary value of the Ernst potential of the transformed solution
\begin{equation}\label{HBT}
\widetilde{\mathcal{E}}_0+\overline{\widetilde{\mathcal{E}}}_0= k_1\dfrac{(\mathcal{E}_0+\overline{\mathcal{E}}_0)(w_0-\beta)} {(\mathcal{E}_0+i k_0)(\overline{\mathcal{E}}_0-i k_0)},\qquad
\widetilde{\mathcal{E}}_0=k_1\dfrac{(w_0-\beta)}{(\mathcal{E}_0+i k_0)}    +i k_2
\end{equation}
where $k_1$,$k_2$ and $w_0$ are arbitrary real constants.
To compare these transformations with Belinski and Zakharov soliton generating transformations on the boundary $\alpha=0$ given by  (\ref{EotoE}), we consider a pair of  subsequent transformations (\ref{HBT}). Below the constants $w_0$, $k_0$, $k_1$, $k_2$ correspond to the first transformation and $\widetilde{w}_0$, $\widetilde{k}_0$, $\widetilde{k}_1$, $\widetilde{k}_2$ -- to the second one:
\[\widetilde{\widetilde{\mathcal{E}}}_0=\dfrac{\widetilde{k}_1 (\widetilde{w}_0-\beta)}{(\widetilde{\mathcal{E}}_0+i \widetilde{k}_0)}
+i \widetilde{k}_2
=\dfrac{\widetilde{k}_1(\widetilde{w}_0-\beta)}    {\dfrac{k_1(w_0-\beta)}{\mathcal{E}_0+i k_0}+i k_2+i\widetilde{k}_0}+i \widetilde{k}_2
\]
This transformation of the boundary values for the Ernst potential coincides with that for Belinski and Zakharov two-soliton transformations described by (\ref{EotoE}), if we choose the following relations between the constants:
\[\begin{array}{l}
w_0=w_1,\\
\widetilde{w}_0=w_2,
\end{array}\qquad
\widetilde{k}_1 =k_1=-\dfrac{(c_1-c_2)(\widetilde{k}_0+k_2)} {c_1 c_2(w_1-w_2)}, \qquad \widetilde{k}_2 =-k_0=\dfrac{1}{c_1}.
\]
Thus, a pair of subsequent Harrison's B$\ddot{a}$cklund transformations with different values of integration constants is equivalent to generation of a pair of Belinski - Zakharov solitons. We recall here that such equivalence of B$\ddot{a}$cklund transformations and the soliton generating technique was affirmed earlier by Cosgrove \cite{Cosgrove:1980} using more complicate and less explicit considerations.

\vspace{-1ex}
\paragraph{\bf\textit{\underline{Neugebauer B$\ddot{a}$cklund transformations for vacuum fields.}}}\quad\hfill\\[1ex]
Another form of B$\ddot{a}$cklund transformations for vacuum fields was presented by Neugebauer \cite{Neugebauer:1979}--\cite{Neugebauer:1980b}, who restricted all his considerations by the stationary axisymmetric fields only. These transformations also lead from any chosen beginning solution to some new family of solutions.
In our notations, for stationary axisymmetric fields we should put $\epsilon=-1$, $\alpha=\rho$ and $\beta=z$. The expression for the Ernst potential $\mathcal{E}(\rho,z)$ for solutions derived after a series of subsequent B$\ddot{a}$cklund transformations of the beginning solution ${\overset \circ{\mathcal{E}}}(\rho,z)$ was presented in \cite{Neugebauer:1979}--\cite{Neugebauer:1980b} in the form (in which we changed a bit the notations):
\begin{equation}\label{Backlund}
\left\vert\begin{array}{ccccc}
{\mathcal{E}}-{\overset \circ{\mathcal{E}}}& 1&1& \cdots & 1\\
{\mathcal{E}}+\overline{{\overset \circ{\mathcal{E}}}}&a_1\gamma_1&a_2\gamma_2& \cdots & a_N\gamma_N\\
{\mathcal{E}}-{\overset \circ{\mathcal{E}}}&\gamma_1^2&\gamma_2^2&\cdots & \gamma_N^2\\
{\mathcal{E}}+\overline{{\overset \circ{\mathcal{E}}}}&a_1\gamma_1^3&a_2\gamma_2^3 &\cdots & a_N\gamma_N^3\\
\vdots&\vdots&\vdots&\cdots & \vdots\\
{\mathcal{E}}-{\overset \circ{\mathcal{E}}}&\gamma_1^N&\lambda_2^N&\cdots & \gamma_N^N
\end{array}\right\vert=0,
\end{equation}
where $N$ should be even; the functions $\gamma_1,\gamma_2,\ldots,\gamma_N$ and $a_1,a_2,\ldots,a_N$  are the values at some chosen points $w_1,w_2,\ldots,w_N$ of two auxiliary functions of $\rho$, $z$ and of a constant complex parameter $w$, such that
\[\gamma\equiv \gamma(w,\zeta,\overline{\zeta})=\left(\dfrac{w-i\overline{\zeta}}{w+i\zeta}
\right)^{\!\! 1/2}\hskip1ex \text{and}\hskip2ex a\equiv a(w,\zeta,\overline{\zeta}),\quad \text{where}\quad
\left\{\!\begin{array}{l}\zeta=\rho+i z,\\[0.5ex]
{\overline{\zeta}}=\rho-i z
\end{array}\right.
\]
The constants $w_1,w_2,\ldots,w_N$ can be chosen arbitrarily, provided only that some of these may be real but if not all, then the set of others should consist of pairs of complex conjugated to each other constants. For real $w_k$, the values $\gamma_k\equiv\gamma(w_k,\zeta,\overline{\zeta})$ should satisfy the condition $\gamma_k\overline{\gamma_k}=1$, but for complex conjugated pair  $w_i$ and $w_k$ the corresponding values of $\gamma$ should be chosen so that  $\gamma_i\overline{\gamma_k}=1$. All $a_k=a(w_k,\zeta,\overline{\zeta})$ are the solutions of the same system of Riccati equations but with different (for different $k$) integration constants:
\begin{equation}\label{a-equation}
\left\{\begin{array}{l}
\dfrac{\partial a_k}{\partial\zeta}=({\overset \circ{\mathcal{E}}}+\overline{{\overset \circ{\mathcal{E}}}})^{-1}
\left[(a_k-\gamma_k)\dfrac{\partial}{\partial\zeta}\overline{{\overset \circ{\mathcal{E}}}}+a_k(a_k\gamma_k-1)\dfrac{\partial}{\partial\zeta}{\overset \circ{\mathcal{E}}}\right],\\[3ex]
\dfrac{\partial a_k}{\partial\overline{\zeta}}=({\overset \circ{\mathcal{E}}}+\overline{{\overset \circ{\mathcal{E}}}})^{-1}
\left[(a_k-\dfrac 1{\gamma_k})\dfrac{\partial}{\partial\overline{\zeta}}\overline{{\overset \circ{\mathcal{E}}}}+a_k(\dfrac{a_k}{\gamma_k}-1)\dfrac{\partial}{\partial\overline{\zeta}}{\overset \circ{\mathcal{E}}}\right],
\end{array}\right.
\end{equation}
where the integration constants should be chosen so that for real $w_k$ the solution $a_k$ satisfies the condition $a_k\overline{a_k}=1$, and for complex conjugated $w_i$ and $w_k$ the corresponding solutions must satisfy the condition $a_i\overline{a_k}=1$.

The equations (\ref{a-equation}), as well as the equations (\ref{Mk-eqs}) are not necessarily can be solved explicitly for any chosen beginning solution $\overset \circ{\mathcal{E}}(\zeta,\overline{\zeta})$. Therefore, if we would not like to make further restrictions on the choice of the beginning (starting) solution, we have to continue our consideration, similarly to the case of generating of solitons considered above, using an asymptotical representation of these B$\ddot{a}$cklund transformations and restricting these by the simplest case $N=2$. For this, we represent the solution (\ref{a-equation}) near $\rho=0$ by a series $a=a_0(z)+\rho^2 a_1(z)+\ldots$ and, taking into account the expansions (\ref{Ernst-expansion}) restricted to a vacuum case, for the leading term $a_0$  we obtain the equation
\[\dfrac{\partial}{\partial z} a_0(z) = \dfrac{a_0(z)-1}{\overset \circ{\mathcal{E}}(z)+\overline{\overset \circ{\mathcal{E}}}(z)} \left[\dfrac{\partial}{\partial z}\overline{\overset \circ{\mathcal{E}}}(z)+ a_0(z)\dfrac{\partial}{\partial z}{\overset \circ{\mathcal{E}}}(z)\right],\quad \overline{a}{}_0(z)=a_0^{-1}(z),
\]
where ${\overset \circ{\mathcal{E}}}(z)$ is the value of the Ernst potential of the beginning solution on the boundary $\rho=0$. This equation admits an explicit solution
\[a_0(z)=(i\ell+\overline{\overset \circ{\mathcal{E}}})/(i\ell-{\overset \circ{\mathcal{E}}}),\]
where $\ell$ is an arbitrary real integration constant. Choosing for $\ell$ subsequently two real values $\ell_1$ and $\ell_2$ and using for $\lambda_k$ the expansion  for $\rho\to 0$ of the form $\lambda_k=1-i\rho/(z-w_k)+\ldots$, we obtain from (\ref{Backlund}) for real $w_k$  the relation between the boundary values of the Ernst potentials corresponding to the B$\ddot{a}$cklund transformations with $N=2$:
\begin{equation}
\mathcal{E}_{\scriptscriptstyle{BT}}(z)=\dfrac{i\left[\ell_2(z-w_2) - \ell_1(z-w_1)\right]
{\overset \circ{\mathcal{E}}}(z)+ \ell_1\ell_2(w_1-w_2)} {(w_1-w_2) {\overset \circ{\mathcal{E}}}(z)+i [\ell_2(z-w_1) - \ell_1(z-w_2)]}
\end{equation}
It is easy to see that this transformation of the boundary values of vacuum Ernst potentials coincides with the two-soliton transformation (\ref{EotoE}), with real $w_1$ and $w_2$, if we choose there ${c_1}=1/{\ell_1}$, ${c_2}=1/{\ell_2}$. It can be shown also that a subsequent B$\ddot{a}$cklund transformations which lead to (\ref{Backlund}), are equivalent to generating vacuum N-soliton solution on the same beginning (background) solution. Thus,
within the class of stationary axisymmetric vacuum gravitational fields with a regular behaviour near some part of the axis  $\rho=0$, for any choice of the beginning (background) solution, the B$\ddot{a}$cklund transformations described in \cite{Neugebauer:1979}  lead to the same transformation of the space of solutions as generation of solitons found earlier in \cite{Belinski-Zakharov:1978,Belinski-Zakharov:1979} and \cite{Harrison:1978} and restricted here for comparison by stationary axisymmetric fields.

\subsubsection*{Exponentiating of Kinnersley-Chitre algebra of symmetries\footnote{It is necessary to mention here, that our notations introduced in the previous part of this paper, differ in some points from Kinnersley and Chitre notations. In particular, we use the metrics of the signuture $(-,+,+,+)$ instead of $(+,-,-,-)$ used in their papers. Our numeration of coordinates and notations for indices defined in (\ref{Components}) -- (\ref{Orbitmetric}) are also different from \cite{KCII:1977} -- \cite{KCIV:1978}. As a result, for metrics on the orbits, we use $h_{ab}$ so that $f_{AB}\to -h_{ab}$. However, in this section, in contrast with other parts of the present paper, we use ${}^\ast$ denoting complex conjugation, the gradient operator $\nabla$ instead of the operator $\partial_\mu$ and the dual operator $\widetilde{\nabla}$ instead of our usual $\varepsilon_\mu{}^\nu\partial_\nu$. Besides that, for stationary axisymmetric fields (which are considered by Kinnersley and Chitre only) we should put in our previous expressions $\epsilon=\epsilon_0=-1$ and $\epsilon_1=\epsilon_2=1$.}}
As it was mentioned in the Introduction, in the papers \cite{Kinnersley:1977}, \cite{KCII:1977} - \cite{KCIV:1978}, Kinnersley and Chitre constructed a representation of an infinite-dimensional algebra of infinitesimal symmetries of Einstein - Maxwell equations for stationary axisymmetric fields using an infinite hierarchies of fields and potentials associated with every solution.
They found also that for vacuum fields these hierarchies admit generating functions determined by the solutions of a linear system with free complex parameter.
Some special kinds of Kinnersley and Chitre infinitesimal symmetry transformations were exponentiated by Hoenselaers, Kinnersley and Xanthopoulos in \cite{HKX:1979a, HKX:1979b}.

\vspace{-1ex}
\paragraph{\bf\textit{\underline{Infinite hierarchies of potentials and their generating functions.}}}\hfill\\[0ex]
The hierarchies of fields  $\{{\overset {n}{\mathcal{H}}}{}_{ab}\}$ and $\{{\overset {n}{\Phi}}{}_{a}\}$ ($a,b=3,4$; $m,n=1,2,\ldots$), associated with any given solution  satisfy the same field equations (\ref{Kinnersley})
\[
\nabla {\overset {n}{\mathcal{H}}}{}_{ab}=i\rho^{-1} h_a{}^c \widetilde{\nabla}{\overset {n}{\mathcal{H}}}{}_{cb},\qquad
\nabla {\overset {n}{\Phi}}{}_{a}=i\rho^{-1} h_a{}^c \widetilde{\nabla}{\overset {n}{\Phi}}{}_{c},
\]
and the fields of these hierarchies are defined recurrently by the relations:
\begin{equation}\label{KCPotentials}
\begin{array}{lcl}
{\overset {\scriptstyle{n+1}}{\Phi}_a}=i({\overset {\scriptstyle{1 n}}{M}_a}+2 \Phi_a{\overset {\scriptstyle{1 n}}{K}}+\mathcal{H}_{ac}{\overset {\scriptstyle{n}}{\Phi^c}}),&& {\overset {\scriptstyle{1}}{\Phi}_a}=\Phi_a,\\
{\overset {\scriptstyle{n+1}}{\mathcal{H}}_{ab}}=i({\overset {\scriptstyle{1 n}}{N}_{ab}}+2 \Phi_a{\overset {\scriptstyle{1 n}}{L}}_b+\mathcal{H}_{ac}{\overset {\scriptstyle{n}}{\mathcal{H}^c{}_b}}),&& {\overset {\scriptstyle{1}}{\mathcal{H}}_{ab}}=\mathcal{H}_{ab},
\end{array}
\end{equation}
where hierarchies of potentials ${\overset {\scriptstyle{m n}}{K}}$, ${\overset {\scriptstyle{m n}}{L}}_b$, ${\overset {\scriptstyle{m n}}{M}}_a$ and ${\overset {\scriptstyle{m n}}{N}}_{a b}$ are defined by equations
\[\nabla{\overset {\scriptstyle{m n}}{K}}={\overset {\scriptstyle{m}}{\Phi}}{}_c^\ast \nabla {\overset {\scriptstyle{n}}{\Phi}}{}^c,\quad
\nabla{\overset {\scriptstyle{m n}}{L}}_b={\overset {\scriptstyle{m}}{\Phi}}{}_c^\ast \nabla {\overset {\scriptstyle{n}}{\mathcal{H}}}{}^c{}_b,\quad
\nabla{\overset {\scriptstyle{m n}}{M}}_a={\overset {\scriptstyle{m}}{\mathcal{H}}}{}_{ca}^\ast \nabla {\overset {\scriptstyle{n}}{\Phi}}{}^c,\quad
\nabla{\overset {\scriptstyle{m n}}{N}}_{ab}={\overset {\scriptstyle{m}}{\mathcal{H}}}{}_{ca}^\ast \nabla {\overset {\scriptstyle{n}}{\mathcal{H}}}{}^c{}_b
\]
in which ${}^\ast$ means complex conjugation, $\nabla$ is a gradient operator such that in Weyl coodinates it is $\nabla=\{\partial_\rho,\partial_z\}$ and
the dual operator $\widetilde{\nabla}=\{\partial_z,-\partial_\rho\}$. Also we have $\mathcal{H}_{ab}=-H_{ab}$, while $2\times 2$-matrix potential $H_{ab}$ and complex electromagnetic potential $\Phi_a$ are defined in (\ref{Kinnersley}).
The action of generators of infinitesimal symmetries on these fields and potentials was described in \cite{KCIII:1978}

\vspace{-1ex}
\paragraph{\rm\textit{\underline{Generating functions for hierarchies of fields and potentials in vacuum.}}}\hfill\\[0ex]
The next very interesting step was made when Kinnersley and Chitre introduced in \cite{KCIII:1978,KCIV:1978} for calculation of the hierarchies of fields ${\overset {n}{\mathcal{H}}}{}_{ab}$ and potentials ${\overset {\scriptstyle{m n}}{N}}_{a b}$ for vacuum, two $2\times 2$-matrix generating functions
\[\left\{\begin{array}{l}
F_{ab}(t)=\mathop\sum_{n=0}^{\infty} t^n {\overset {n}{\mathcal{H}}}{}_{ab},\\[1ex]
F_{ab}(0)=i \epsilon_{ab},
\end{array}\right.\qquad
\left\{\begin{array}{l}
G_{ab}(s,t)=\mathop\sum_{m,n=0}^{\infty} s^m t^n {\overset {m n}{N}}{}_{ab},\\[1ex]
G_{ab}(0,t)=-i F_{ab}(t).
\end{array}\right.
\]
where $t$ and $s$ are auxiliary complex parameters and coordinate dependence of the generatin functions and coefficients is omitted. A beautiful discovery in
\cite{KCIII:1978,KCIV:1978} was that $F_{ab}(t)$ should satisfy a matrix linear equation and the function $G_{ab}$ possesses very simple expression in terms of $F_{ab}$:
\begin{equation}\label{FGmatrices}
\begin{array}{lcl}
\nabla F_{ab}=i t S^{-2}\left[(1-2 t z)\nabla H_{ac}-2 t\rho \widetilde{\nabla} H_{ac}\right] F^c{}_b,&& F_{ab}(0)=i \epsilon_{ab},\\[2ex]
G_{ab}(s,t)=(s-t)^{-1}\left[s \epsilon_{ab}+t S(s) F_{ca}(s)F^c{}_b(t)\right],
&& G_{ab}(0,t)=-i F_{ab}(t),
\end{array}
\end{equation}
where the function $S^2(t)=(1-2 t z)^2+(2 t\rho)^2$.

\vspace{-1ex}
\paragraph{\bf\textit{\underline{``HKX'' vacuum-to-vacuum rank $p\ge 0$ transformations.}}}\hfill\\[1ex]
Using the generating functions $F_{ab}$ and $G_{ab}$ for the hierarchies of fields and potentials, C.Hoenselaers, W.Kinnersley and B.C.Xanthopoulos \cite{HKX:1979a,HKX:1979b} have been able to exponentiate some special kinds of Kinnersley-Chitre ``vacuum-to-vacuum'' infinitesimal symmetry transformations and obtain a series of finite symmetry transformations of different ``ranks''. Namely, for the simplest case of rank $p=0$  transformations they obtain for the Ernst potential
\begin{equation}\label{HKX0}
\mathcal{E}={\overset o {\mathcal{E}}}+\dfrac{i\alpha G(0,u)}{1-\alpha G(u,u)}\left[\vphantom{\nabla^x_y}\partial_t G(u,t)\right]_{t=0}
\end{equation}
where $\alpha$ and $u$ are arbitrary real constants and  $G(s,t)$ is the upper left element of the matrix $G_{ab}(s,t)$ defined in (\ref{FGmatrices}).

For transformations of rank $p \ge 1$ the   expressions are more complicate:
\begin{equation}\label{HKXp}
\mathcal{E}={\overset o {\mathcal{E}}}+i\alpha^{(p)}\mathop\sum_{k,l=0}^{p} G_{0,p-k}(0,u)M_{kl}^{-1}(u)\left[\vphantom{\nabla^x_y}\partial_t G_{l,0}(u,t)\right]_{t=0},
\end{equation}
where the elements of the matrices $M_{ik}(u)$ and $G_{ij}(s,t)$ are defined as follows:
\begin{equation}\label{Mmatrix}
M_{ik}(u)=\delta_{ik}-\alpha^{(p)} G_{i,p-k}(u,u),\quad
G_{ij}(s,t)=\dfrac{s^i\, t^j}{i!\,j!}\left(\dfrac{\partial}{\partial s}\right)^i
\left(\dfrac{\partial}{\partial t}\right)^j G(s,t)
\end{equation}
and the parameters $\alpha^{(p)}$ and $u$ are arbitrary real constants.

In the most general case, the authors of \cite{HKX:1979b} suggested to consider a combined transformations which infinitesimal versions can include a finite sums of transformations of different ranks which can possess different values of the parameter $u$. However, below we consider only the transformations (\ref{HKX0}) and (\ref{HKXp}), but our analysis can be applied to more general cases as well.

\vspace{-1ex}
\paragraph{\bf\textit{\underline{HKX-transformations  of the axis values of the Ernst potentials.}}}\hfill\\[1ex]
Similarly to our considerations of other solution generating methods presented above, we consider in this subsection the transformations of the axis data for the Ernst potentials corresponding to HKX-transformations.

\vspace{-1ex}
\paragraph{\rm\textit{\underline{Generating functions for hierarchies of  potentials on the axis of symmetry.}}}\hfill\\[0ex]
Using the asymptotic behaviour of metric components and Ernst potentials (\ref{HOmegaf-expansion}) and (\ref{Ernst-expansion}) near the regular parts of the axis of symmetry, where for stationary axisymmetric vacuum fields we should set $\alpha=\rho$, $\beta=z$, and considering all electromagnetic components as vanishing, we can solve the equations (\ref{FGmatrices}) asymptotically and on the axis $\rho=0$  we obtain
\begin{equation}\label{FGAxis}
F_{ab}(t)=\begin{pmatrix}
\dfrac{t {\overset o {\mathcal{E}}}(z)}{1-2 t z}& \dfrac{i}{1-2 t z}\\[2ex]
-i& 0
\end{pmatrix},\qquad    G_{ab}(s,t)=-i F_{ab}(t),
\end{equation}
i.e., the matrix $G_{ab}$ on the axis $\rho=0$ does not depend on $s$. This is a great simplification for calculations of HKX-transformations of the axis data.

\vspace{-1ex}
\paragraph{\rm\textit{\underline{Rank-0 HKX-transformations in terms of Ernst potentials on the axis.}}}\hfill\\[0ex]
From the general form of rank-0 transformations \cite{HKX:1979b} shown in (\ref{HKX0})\footnote{We recall here that in (\ref{HKX0}), $G$ means the upper left component of the matrix $G_{ab}$.}, and the derived above expressions for generating functions on the axis of symmetry (\ref{FGAxis}), we obtain for the generating vacuum solution on the axis
\begin{equation}\label{HKX0Axis}
\mathcal{E}(z)=\dfrac{(w_0-z){\overset o {\mathcal{E}}}(z)}{
w_0-z+i(\alpha/2)\, {\overset o {\mathcal{E}}}(z)}
\end{equation}
for any ``seed'' vacuum solution characterized on the axis by the Ernst potential ${\overset o {\mathcal{E}}}(z)$. In (\ref{HKX0Axis}) we put $u=1/(2 w_0)$. The parameter $w_0$ can be eliminated after a shift of the coordinate $z$ along the axis: $z\to z+w_0$. Thus, this transformation depends on one essential real parameter $\alpha$ only. From physical point of view, this transformation leads to generation of the solution corresponding to superposition of the background field of the ``seed'' solution and the field of extreme Kerr source restricted by a strange subcase with zero mass and with angular momentum and NUT parameters $a=b=-\alpha/4$. This is a very restricted subcase of vacuum two-soliton solutions (\ref{EotoE}).

\vspace{-1ex}
\paragraph{\rm\textit{\underline{Combined HKX-transformations in terms of Ernst potentials on the axis.}}}\hfill\\[0ex]
Following the set of examples considered in \cite{HKX:1979b} (applied there to the flat space-time only), we consider here the transformation, combined of two rank-0 ones applied, however, to arbitrary ``seed'' solution with the Ernst potential ${\overset o {\mathcal{E}}}(z)$ on the axis. In this case, for the transformed solution we have
\begin{equation}\label{HKX0x2Axis}
\mathcal{E}(z)=\dfrac{(w_1-z)(w_2-z){\overset o {\mathcal{E}}}(z)}{
(w_1-z)(w_2-z)+(i/2)[\alpha_1(w_2-z)+\alpha_2(w_1-z)]{\overset o {\mathcal{E}}}(z)}
\end{equation}
where $w_1=1/(2 u_1)$, $w_2=1/(2 u_2)$ and $\alpha_1$, $\alpha_2$ are four arbitrary real constants. One of constants $w_1$, $w_2$  or their combination can be eliminated by a shift of the coordinate $z$ along the axis. This transformation generates the fields of two extreme objects of the type (\ref{HKX0}) interacting with each other and with the ``seed'' metric characterised by the Ernst potential ${\overset o {\mathcal{E}}}(z)$ on the axis. This is also very restricted  subcase of vacuum 4-soliton solutions.

\vspace{-1ex}
\paragraph{\rm\textit{\underline{Rank-p ($p\ge 1$) HKX-transformations of the Ernst potential on the axis.}}}\hfill\\[1ex]
The rank-$p$ HKX-transformations of the Ernst potentials were defined by the expressions (\ref{HKXp}). Calculating these transformations on the axis of symmetry, we have to take into account, that in the sum (\ref{HKXp}) the only nonvanishing term corresponds to the indices $k=l=0$. To explain this, we note, that in (\ref{HKXp}), in each product in the sum the last multiplier is nonvanishing on the axis only for $l=0$ because on the axis $G(s,t)$ is independent of $s$ due to (\ref{FGAxis}). Also, it is easy to see that because of the same reason, the matrix $M_{ik}$ on the axis is upper triangular. Therefore in this matrix and in its inverse, the  first column possess only one component which is nonvanishing on the axis. This is $M_{00}$. The same is true for the matrix $M_{ik}^{-1}$ and its upper left element on the axis is $M^{-1}{}_{00}=1/M_{00}$. Therefore, we obtain from (\ref{HKXp}) on the axis:
\begin{equation}\label{HKXpAxis}
\mathcal{E}(z)=\dfrac{(w_0-z)^{p+1}{\overset o {\mathcal{E}}}(z)}{
(w_0-z)^{p+1}+i(\alpha^{(p)}/2) w_0 z^{p-1}\, {\overset o {\mathcal{E}}}(z)}
\end{equation}
The rank-$p$  transformation depends on two essential real parameter $\alpha$, and $w_0$. In this case, the parameter $w_0=1/(2 u)$, can not be eliminated by a shift of the origin of the coordinate $z$ along the axis. Nonetheless,  an absence of large enough number of free parameters do not allow these solutions to have a rich physical interpretation like in the soliton generating cases. One can note also that if the transforming solution is chosen asymptotically flat, the expansion of the tranformed Ernst potential on the axis (\ref{HKXpAxis}) for $z\to\infty$ allows to determine the changes in the multipole moments of the transformed solution with respect to the transforming one. In particular, it is easy to see, that the transformation (\ref{HKXpAxis}) do not change the parameter of mass, but not the same is for the NUT parameter. It is clear that for generating physically more interesting solutions by this method, one should consider the combined HKX-transformations of different ranks and with different values of the parameter $w\equiv 1/(2 u)$. However, the constructions of such transformations are much more complicate and these had not been considered by these authors.

\subsubsection*{Hauser and Ernst ``effectivization'' of infinitesimal Kinnersley-\\Chitre transformations}
A powerful approach to exponentiation (``effectivization'') of Kinnersley-Chitre algebra of infinitesimal symmetries of stationary axisymmetric Einstein-Maxwell equations was developed by I.Hauser and F.J.Ernst in \cite{HE:1979a} --\cite{HE:1980d}.

\vspace{-2ex}
\paragraph{\rm\textit{\underline{Basic assumptions.}}} In Hauser-Ernst  approach, all stationary axisymmetric solutions  are assumed to be regular in some neighbourhood of at least one point on the symmetry axis.

\vspace{-2ex}
\paragraph{\rm\textit{\underline{Hauser-Ernst approach to exponentiation of vacuum symmetries.}}}  Within\\ this class of vacuum fields, in \cite{HE:1979a} for Kinnersley-Chitre matrix potentials $\mathbf{F}(t)$ depending on Weyl coordinates $\rho$,$z$ and on a free complex parameter $t$ (here and below the dependence on Weyl coordinates is omitted), a $2\times 2$-matrix homogeneous Hilbert problem (HHP) was formulated and $2\times 2$-matrix linear singular integral equation solving this problem was derived.

\vspace{-2ex}
\paragraph{\rm\textit{\underline{$3\times 3$-matrix $\mathbf{F}$-potential for electrovacuum fields.}}}
In \cite{HE:1979b}, this approach was generalized to the similar class of electrovacuum fields. Instad of the Kinnersley-Chitre vacuum matrix potential, these authors constructed a $3\times 3$-matrix potential $F(t)$ which satisfies to generalized linear system with a complex parameter $t$. It was argued also that for the fields, which are regular on some part of the axis, the gauge can be chosen so that $\mathbf{F}(t)$ is holomorphic for all $t$ besides two branching points $2t=1/(z+i\rho)$ and $2t=1/(z-i\rho)$ and $t=\infty$, but $\mathbf{F}(t)\cdot\text{diag}\{t,1,1\}$ is holomorphic at $t=\infty$. Besides that, such $\mathbf{F}$-potentials can be chosen so that $\mathbf{F}(0)=\left(\begin{smallmatrix} 0&i&0\\ -i&0&0\\ 0&0&1\end{smallmatrix}\right)$.

\vspace{-2ex}
\paragraph{\rm\textit{\underline{Hauser-Ernst homogeneous Hilbert problem.}}}
The homogeneous Hilbert problem was formulated in \cite{HE:1979b}, \cite{HE:1980c} on a closed contour $L$ on $t$-plane, which is symmetric with respect to a real axis and surrounding $ t=0$, so that the mentioned above branch points $2t=1/(z+i\rho)$ and $2t=1/(z-i\rho)$ are outside this contour. Then, the homogeneous Hilbert problem looks as
\begin{equation}\label{HE-HHP}
\mathbf{X}_-(t)=\mathbf{X}_+(t) \mathbf{G}(t),\quad t\in L,\quad \mathbf{X}_+(0)=\mathbf{I},
\end{equation}
where $\mathbf{X}_+(t)$ is holomorphic on $\mathcal{L}_+$ and $\mathbf{X}_-(t)$ -- on $\mathcal{L}_-$ (here $\mathcal{L}_+$ means the region
on and inside $L$, while $\mathcal{L}_-$ means the region
on and outside $L$ including $t=\infty$). The matrix $\mathbf{G}(t)$ connects $\mathbf{X}_+(t)$ and $\mathbf{X}_-(t)$ on the contour $L$.

\vspace{-2ex}
\paragraph{\rm\textit{\underline{$\mathbf{F}$-potentials from the
solution homogeneous Hilbert problem.}}}
Various $\mathbf{F}$-po\-tentials  are related to solutions of the homogeneous Hilbert problem (\ref{HE-HHP}) as
\[\left\{\begin{array}{ll}
\mathbf{F}(t)=X_+(t)\cdot {\overset o {\mathbf{F}}}(t),& t\in \mathcal{L}_+\\
\mathbf{F}(t)\cdot \mathbf{u}(t)=X_-(t)\cdot {\overset o {\mathbf{F}}}(t),&
t\in \mathcal{L}_-
\end{array}\right.
 \qquad\mathbf{G}(t)={\overset o {\mathbf{F}}}(t)\cdot\mathbf{u}(t)\cdot {\overset o {\mathbf{F}}}^{-1}(t) \]
where ${\overset o {\mathbf{F}}}(t)$ is the $\mathbf{F}$-potential of the initial (``seed'') solution. The matrix $\mathbf{u}(t)$ should satisfy the algebraic conditions:
\begin{equation}\label{u-matrix}
\left\{\begin{array}{l}
\mathbf{u}^\dagger(t)\cdot\mathcal{G}\cdot \mathbf{u}(t)=\mathcal{G},\\[1ex]
\det\mathbf{u}(t)=1,
\end{array}\right.\qquad \mathcal{G}=\begin{pmatrix}
0&1&0\\
-1&0&0\\
0&0&t/(2 i)
\end{pmatrix},
\end{equation}
Besides that, $\mathbf{u}(t)$ should be holomorphic in $\mathcal{L}_-$, and the products with its components $t u^2{}_1$, $t u^3{}_1$, $t^{-1} u^1{}_2$, $t^{-1} u^1{}_3$  should be holomorphic at $t=\infty$.
\vspace{-2ex}
\paragraph{\rm\textit{\underline{Linear singular integral equation solving the homogeneous Hilbert problem.}}}\hfill\\[0ex]
In \cite{HE:1979b}, \cite{HE:1980c} a $3\times 3$-matrix linear singular integral equation, solving the (\ref{HE-HHP}) was derived as the condition of holomorphicity of $\mathbf{X}_-(t)$ in $\mathcal{L}_-$:
\begin{equation}\label{HE_equation}
\int_L\dfrac{\mathbf{F}(s)\cdot \mathbf{u}(s)\cdot{\overset o {\mathbf{F}}}{}^{-1}(s)}
{s(s-t)}\, ds=0.
\end{equation}
Thus, I.Hauser and F.J.Ernst constructed a solution generating method in which for every choice of the seed  (transforming) solution potential ${\overset o {\mathbf{F}}}(t)$ and every chosen $\mathbf{u}(t)$ with the analytical properties described above the $\mathbf{F}$-potential can be determined from the integral equation (\ref{HE-equation}). In this method, the selection of $3\times 3$-matrix $\mathbf{u}(t)$ corresponds to some particular  element of Kinnersley-Chitre algebra of infinitesimal symmetry transformations.

\vspace{-2ex}
\paragraph{\rm\textit{\underline{Hauser-Ernst solution generating method as transformations of axis data.}}}\hfill\\[0ex]
In Hauser-Ernst gauges, the value of $\mathbf{F}$-potentials on the axis was chosen as:
\begin{equation}\label{HE_Faxis}
\mathbf{F}(t)=\begin{pmatrix}
0&i&0\\
-\dfrac{i}{1-2 t z}& \dfrac{t {\mathcal{E}}_o(z)}{1-2 t z}&
\dfrac{t{\Phi}_o(z)}{1-2 t z}\\
0&0&1
\end{pmatrix}
\end{equation}
As it was shown in \cite{HE:1980c}, the equation (\ref{HE_equation}), considered on the axis, implies
\begin{equation}\label{HE-CuD}\mathbf{C}(t)\cdot \mathbf{u}(t)\cdot\mathbf{D}(t)=0 \quad\left\Vert\quad
\begin{array}{l}
\mathbf{C}(t)=\{-i,\,t{\mathcal{E}}(1/2 t),\,t{\Phi}(1/2 t)\}\\[1ex]
\mathbf{D}(t)=\left(\begin{array}{ll}
-t{\overset o{\mathcal{E}}}(1/2 t)&-i t{\overset o {\Phi}}(1/2 t)\\
-i& 0\\
0& 1
\end{array}\right)
\end{array}\right.
\end{equation}
where ${\overset o{\mathcal{E}}}(z)$, ${\overset o {\Phi}}$ are the Ernst potentials of the chosen initial (seed) solution and  $\mathcal{E}(z)$, $\Phi(z)$ are the Ernst potential for the generated (transformed) solution, all considered on the axis $\rho =0$ and with substitution $z\to 1/(2 t)$.

To obtain explicitly the transformation of the axis data for the Ernst potentials corresponding to Hauser-Ernst solution generating method, we note that the condition (\ref{HE-CuD}) can be solved explicitly and thus we obtain
\begin{equation}\label{HE-EoFotoEF}\left.\begin{array}{l}
\mathcal{E}(z)=-2 i z\dfrac{\mathcal{G}(z)}{\mathcal{D}(z)}\\
\Phi(z)=-2 i z\dfrac{\mathcal{F}(z)}{\mathcal{D}(z)}
\end{array}\hskip1ex\right\Vert\hskip1ex \{\mathcal{D}(z),\mathcal{G}(z),\mathcal{F}(z)\}=\{2 i z, -{\overset o{\mathcal{E}}}(z),-{\overset o{\Phi}}(z)\}\cdot \mathbf{u}^{-1}(\dfrac{1}{2 z})
\end{equation}
where in $\mathbf{u}(t)$,  we made a substitution $t\to 1/(2 z)$. However, it is clear that in these expressions, we can not choose the components of $\mathbf{u}(t)$ arbitrarily because these should satisfy the certain restrictions (see (\ref{u-matrix}) and two lines after it). To solve these conditions we use a Gauss decomposition for $\mathbf{u}(t)$:
\begin{equation}\label{Gauss}\mathbf{u}(t)=\begin{pmatrix}
1&0&0\\
p_1&1&0\\
p_2&p_3&1
\end{pmatrix}\cdot
\begin{pmatrix}
d_1&0&0\\
0&d_2&0\\
0&0&1/(d_1 d_2)
\end{pmatrix}\cdot
\begin{pmatrix}
1&q_1&q_2\\
0&1&q_3\\
0&0&1
\end{pmatrix}
\end{equation}
Here all eight complex parameters are functions of $t$ analytical in $\mathcal{L}_-$. This expression already takes into account that $\det \mathbf{u}(t)=1$. The condition of the behaviour of the components of $\mathbf{u}(t)$ at $t\to\infty$, mentioned after (\ref{u-matrix}), imply that $t\,p_1(t)$, $t\,p_2(t)$, $t^{-1} q_1(t)$, $t^{-1} q_2(t)$ should be nalaytical at $t\to\infty$.

After substitution of (\ref{Gauss}) into the first condition in (\ref{u-matrix}), the corresponding equations can be solved explicitly, but four complex functions of $t$ (or of $1/(2 z)$) remain arbitrary. Therefore, in the expressions (\ref{HE-EoFotoEF}), too many arbitrary functions remain, because to transform the axis data of a seed Ernst potentials ${\overset o {\mathcal{E}}}(z)$, ${\overset o {\Phi}}(z)$ into any arbitrarily chosen axis value of transformed Ernst potentials $\mathcal{E}(z)$, $\Phi(z)$, we need only two analytical complex functions (or four such functions which take real values on the real axis). Thus, we can imose the restrictions on the functional parameters in (\ref{Gauss}) to  exclude pure gauge transformations and to simplify the expressions (\ref{HE-EoFotoEF}). In probably most simple case, we obtain the following (particular) solution of (\ref{u-matrix}):
\[p_1=p_2=q_3=0,\hskip1ex p_3=-4 i z d_1 q_2^\dagger,\hskip1ex d_1=d_1^\dagger,\hskip1ex d_2=d_1^{-1},\hskip1ex q_1-q_1^\dagger=-4 i z q_2 q_2^\dagger
\]
where ${}^\dagger$ means a complex conjugation of a function at complex conjugated point, e.g. $ d_1^\dagger(t)=\overline{d_1(\overline{t})}$.
In this case, using more simple notations, we obtain the axis form of the Hauser-Ernst solution generating transformations:
\begin{equation}\label{HE-Axis}
\left\{\begin{array}{l}
\mathcal{E}(z)=k^2(z)\bigl[{\overset o {\mathcal{E}}}(z)+i \delta(z)-2 c^\dagger(z){\overset o {\Phi}}(z)-c(z) c^\dagger(z)\bigr]\\[1ex]
\Phi(z)=k(z)\bigl[{\overset o {\Phi}}(z)+c(z)\bigr]
\end{array}\right.
\end{equation}
where  $k(z)\equiv d_1(\dfrac 1{2 z})$,\, $\delta(z)\equiv z \bigl[q_1(\dfrac 1{2 z})+q_1^\dagger(\dfrac 1{2 z})\bigr]$,\, $c(z)\equiv 2 i z q_2(\dfrac 1{2 z})$. The functions $k(z)$ and $\delta(z)$ should take the real values for real arguments. In the case ${\overset o {\Phi}}(z)=0$ and $c(z)=0$, (\ref{HE-Axis}) describes vacuum-to-vacuum transform. It is determined (besides the seed  ${\overset o {\mathcal{E}}}(z)$) by two functions
$k(z)$ and $\delta(z)$.

In general, the analytical functions - the components of the matrix $\mathbf{u}(t)$ determined by the functions $k(z)$, $\delta(z)$ and $c(z)$ (together with the seed F-potential matrix) should be used in the kernel of the Hauser-Ernst  integral equation (\ref{HE_equation}) which solution determines the corresponding generated solution outside the axis of symmetry. Thus, at least in principle, the Hauser-Ernst integral equation method allows to generate any solutions from the class of stationary axisymmetric electrovacuum solutions with regular behaviour at least on some part of the axis of symmetry and corresponding to any element of Kinnersley-Chitre algebra of internal symmetries of stationary axisymmetric Einstein - Maxwell equations. However, solution of this integral equation for some non-trivial input data is not a simple task. Only very simple examples with rational input data can be found in the literature and these examples surely are covered by vacuum and electrovacuum soliton generating techniques described earlier in this paper.

\section*{Summary and conclusions}
As it is well known, the $\mathcal{G}_2$-symmetry-reduced vacuum Einstein equations or electrovacuum Einstein-Maxwell equations are integrable and admit various solution generating procedures, which allow to construct large families of exact solutions starting from arbitrarily chosen ``beginning'' (of ``seed'', or ``background'') solution. Each of such solution generating procedures can be considered as transformations of the corresponding solution spaces described in terms of transformations of ``coordinates'' characterizing every local solution.
In the entire solution spaces of the integrable reductions of Einstein's field  equations, the monodromy data for the fundamental solution of associated ``spectral problems'' can be used  as the ``coordinates'' in the infinite-dimensional solution spaces (\cite{Alekseev:1985, Alekseev:1987, Alekseev:2001}).

In this paper, we considered more simple construction of such ``coordinates'' which exist in the (infinite-dimensional) subspaces of electrovacuum solutions for which the gravitational and electromagnetic fields possess a regular behaviour near  degenerate orbits of the space-time isometry group $\mathcal{G}_2$\footnote{Some examples of such subspaces of solutions are cylindrical waves, stationary axisymmetric fields created by compact sources and considered near some intervals on the axis between or outside the sources, some cosmological-like solutions, plane waves near the ``focusing singularities'', solutions with Killing horizons and some others.}. In the corresponding two-dimensional orbit space, the degenerate orbits constitute the lines which can be considered as the boundaries of this orbit space. In the infinite-dimensional spaces of solutions of these types, the values of the Ernst potentials on such boundaries in the orbit spaces can serve as ``coordinates'' of each solution.  The  solution generating transformations considered above was described in terms of these ``coordinates'' by simple expressions in which a particular choice of the beginning (initial, or background) solution is not assumed. The beginning solution in these expressions is represented also by its boundary values of the Ernst potentials which can be chosen as arbitrary functions of the parameter along the boundary. It is clear that many physical parameters of generating solutions can be calculated directly from these boundary values of the Ernst potentials and the detail knowledge of the components of the solution on the whole orbit space is not necessary for these. Besides that, the explicit form of each of these solution generating procedures in terms of these ``coordinates'' allows to compare different suggested solution generating procedures, to find the relations between numerous constant parameters introduced by different methods as well as to determine various physical and geometrical properties of generating solutions  (such as e.g., cylindrical wave profiles on the axis of symmetry, multipole moments of asymptotically flat fields, appearance of horizons in stationary axisymmetric fields and others) even before a detail calculation of all components of generated solution.

It is necessary to note that in this paper we have not discussed the known methods of the other type (also based on the integrability of the equations under consideration) which allow a direct construction of multiparametric families of solutions, but  which do not admit arbitrary choice of some beginning (background) solution. Among these, we can mention the methods for construction of solutions for boundary value problems \cite{Alekseev:1993}, \cite{MAKNP:2008}, for the known algebro-geometric ansatz \cite{Korotkin-Matveev:1990}, for the characteristic initial value problems \cite{Alekseev-Griffiths:2004}.

Besides that, there exist the integral equation methods for a direct construction of solutions. Among these, there is a scalar linear integral equation method which was suggested by N.Sibgatullin \cite{Sibgatullin:1984} for construction of stationary axisymmetric electrovacuum fields. Actually, Sibgatullin  started from the solution generating method suggested earlier by Hauser and Ernst \cite{HE:1979a} -- \cite{HE:1980d} for effecting Kinnersley-Chitre transformations and reduced considerably their matrix linear singular integral equation to a scalar one, using the choice of Minkowski spacetime as the initial solution. During the last decades the Sibgatullin's integral equation  was  used frequently enough in the literature by some authors for construction of particular stationary axisymmetric asymptotically flat solutions with various rational structures of the Ernst potentials on the axis.

Another method of direct construction of solutions of integrable reductions of Einstein's field equations (not only for the case of electrovacuum) which we do not discuss here, is the  monodromy transform approach \cite{Alekseev:1985}, \cite{Alekseev:1987}.  This approach is also based on a reformulation of integrable reductions of Einstein's field equations in terms of a linear singular integral equations, but in contrast to the Hauser and Ernst approach, the construction of these integral equations does not assume any  restrictions on the entire space of local solutions of these field equations. One of general applications of this approach, which can be found in \cite{Alekseev:1992}, is a construction in a unified (determinent) form of a huge class of electrovacuum solutions with \emph{arbitrary} rational structure of the Ernst potentials on degenerate orbits of space-times isometry group $\mathcal{G}_2$.  This class includes hierarchies of soliton and non-soliton solutions, stationary axisymmetric solutions (which are not necessary asymptotically flat), as well as various types of waves and cosmological solutions which admit $\mathcal{G}_2$-symmetry. Using this method, more singular types of solutions for interacting waves and inhomogeneous cosmologies were found in \cite{Alekseev-Griffiths:2000}.

However, a detail consideration and farther comparison of these methods and the results of their applications are not in the scope of this paper.

\subsection*{Acknowledgements}
The work of GAA was supported in parts by the Russian Foundation for Basic Research (grant 18-01-00273 a).

\end{document}